\documentstyle[11pt,aaspp4,eqsecnum]{article}

\def\be{\begin{equation}}
\def\ee{\end{equation}}
\def\ba{\begin{eqnarray}}
\def\ea{\end{eqnarray}}
\def\go{\mathrel{\raise.3ex\hbox{$>$}\mkern-14mu
             \lower0.6ex\hbox{$\sim$}}}
\def\lo{\mathrel{\raise.3ex\hbox{$<$}\mkern-14mu
             \lower0.6ex\hbox{$\sim$}}}
\def\hatl{{\hat l}}
\def\hatmu{{\hat\mu}}
\def\hatx{{\hat x}}
\def\haty{{\hat y}}
\def\hatz{{\hat z}}
\def\hatr{{\hat r}}
\def\hato{{\hat\omega}}
\def\bK{{\bf K}}
\def\bF{{\bf F}}
\def\bN{{\bf N}}
\def\bT{{\bf T}}
\def\br{{\bf r}}
\def\cJ{{\cal J}}
\def\cN{{\cal N}}
\def\bJ{{\bf J}}
\def\bo{{\mbox{\boldmath $\omega$}}}
\def\bcN{{\mbox{\boldmath $\cal N$}}}
\def\bV{{\bf V}}
\def\bB{{\bf B}}

\begin{document}
\title{MAGNETICALLY DRIVEN WARPING, PRECESSION AND RESONANCES
IN ACCRETION DISKS}
%\title{Warping and Precession of Accretion Disks Around Magnetic Stars}

\author{Dong Lai}
\affil{Center for Radiophysics and Space Research, 
Department of Astronomy, Cornell University,
Ithaca, NY 14853;\\
E-mail: dong@spacenet.tn.cornell.edu}
%(Submitted to ApJ, 3/3/99; to appear on the Nov.~1 issue, Vol.~525)}

%%%%%%%%%%%%%%%%%%%%%%%%%%%%%%%%%%%%%%%%%%%%%%%%%%%%%%%%%
\begin{abstract}
The inner region of the accretion disk onto a rotating magnetized
central star (neutron star, white dwarf or T Tauri star) is 
subjected to magnetic torques which induce 
warping and precession of the disk. The origin of these torques lies in
the interaction between the surface current on the disk and the 
horizontal magnetic field (parallel to the disk) produced by the 
inclined magnetic dipole:
The warping torque relies on the radial surface current generated by
the twisting of the vertical field threading the disk, while the 
precessional torque relies on the azimuthal screening current
due to the diamagnetic response of the disk. 
Under quite general conditions, there exists a magnetic
warping instability in which the magnetic torque drives the
disk plane away from the equatorial plane of the star toward 
a state where the disk normal vector is perpendicular to the
spin axis. Viscous stress tends to suppress the warping
instability at large radii, but the magnetic torque always dominates 
as the disk approaches the magnetosphere boundary. 
%This suggests that the inner accretion disk of a magnetized star 
%``prefers'' to be in a state where the disk normal vector is 
%perpendicular to the stellar spin.
The magnetic torque also drives the tilted inner disk 
into retrograde precession (opposite to the rotation of the disk) 
around the stellar spin axis. Moreover, resonant magnetic forcing on 
the disk can occur which may affect the dynamics of the disk. 

The magnetically driven warping instability and precession 
may be related to a number of observational puzzles.
%potentially explain a number of puzzles related to accretion onto
%magnetized stars. 
Examples include:  (1) Spin evolution of 
accreting X-ray pulsars: It is suggested that the observed torque
reversal of the disk-fed magnetized neutron stars is associated 
with the wandering of the inner disk around the preferred perpendicular 
state. (2) Quasi-periodic oscillations in low-mass X-ray binaries: 
The magnetic torque induces disk tilt, making it possible to explain 
the observed low-frequency QPOs using disk precession. 
(3) Super-orbital periods in a number of X-ray binaries as a 
result of warped, precessing disks. (4) Photometric period
variations of T Tauri stars. 
\end{abstract}


%%%%%%%%%%%%%%%%%%%%%%%%%%%%%%%%%%%%%%%%%%%%%%%%%%%%%%%%%%%%%%%%%
\section{INTRODUCTION}

Interaction between an accretion disk and a magnetized central object
lies at the heart of the physics of a variety of astrophysical systems,
including accreting neutron stars, white dwarfs and pre-main-sequence stars
(e.g., Frank et al.~1992; Hartmann 1998). The basic picture of 
disk-magnetosphere interaction was first outlined by Pringle \& Rees (1972)
following the discovery of accretion-powered X-ray pulsars. These are 
rotating, highly magnetized ($B\sim 10^{12}$~G) neutron stars that accrete
material from a companion star, either directly from a stellar wind, or
in the form of an accretion disk. The strong magnetic
field disrupts the accretion flow at the magnetospheric boundary (typically
at a few hundreds neutron star radii),
and channels the plasma onto the polar caps of the neutron star.
The magnetosphere boundary is located where 
the magnetic and plasma stresses balance, 
\be
r_m=\eta \left({\mu^4\over GM\dot M^2}\right)^{1/7},
\label{alfven}\ee
where $M$ and $\mu$ are the mass and magnetic moment of the central object,
$\dot M$ is the mass accretion rate, $\eta$ is a dimensionless constant
of order unity (For an aligned dipole, estimate of $\eta$ ranges from 
$0.5$ to $1$; e.g., Pringle \& Rees 1972; Lamb, Pethick \& Pines 1973;
Ghosh \& Lamb 1979a,b,~1992; Aly 1980; Arons 1993; Wang 1995).
%The star generally gets spun up until its rotation rate 
%equals the Keplerian rotation at the magnetospheric radius; 
%after that point the spin frequency may wonder around the equilibrium 
%value.  This picture can also be applied to weakly magnetized 
%($B\sim 10^8$~G) neutron stars, such as those in low-mass X-ray binaries
In low-mass X-ray binaries containing weakly magnetized 
($B\sim 10^8$~G) neutron stars, the magnetosphere lies close to the stellar
surface; the neutron stars are spun up and eventually become
millisecond pulsars (Alpar et al.~1982; see Phinney \& Kulkarni 1994 or 
Bhattacharya 1995 for a review).  
Similar magnetosphere-disk interaction also occurs in T Tauri stars (e.g.,
Hartmann 1998), where the stellar magnetic field ($B\sim 10^3$~G at the
surface) strongly affects the accretion flow, 
and provides a key ingredient to explain the anomalous slow rotation of
these objects (e.g., K\"onigl 1991; Cameron \& Campbell 1993;
Shu et al.~1994). Finally, models of
magnetized accretion disks have been applied to explain the observed
UV and X-ray emission from some intermediate polars (the DQ Her subclass of
CVs) (e.g., Yi \& Kenyon 1997). 

A large number of theoretical papers have been written on the subject of 
the interaction between accretion disk and a magnetized star
(e.g., Pringle \& Rees 1972; Ghosh \& Lamb 1979a,b; Aly 1980;
Lipunov \& Shakura 1980; Anzer \& B\"orner 1980,1983; Arons 1987,1993; Wang
1987,1995; Aly \& Kuijpers 1990; Spruit \& Taam 1990,1993; Shu et al.~1994; 
van Ballegooijen 1994; Lovelace et al.~1995,1998; 
Li, Wickramasinge \& R\"udiger 1996; Campbell 1997), 
and numerical study of this problem is still in its infancy
(e.g., Stone \& Norman 1994; Hayashi et al.~1996; Goodson et al.~1997;
Miller \& Stone 1997; see also Toropin et al.~1999).
Beyond the simple notion of magnetosphere (with the Alf\'ven radius 
estimated by eq.~[\ref{alfven}]), 
there is little consensus on the range and strength 
of magnetic interaction outside the magnetosphere boundary,
on the efficiency of magnetic field dissipation in/outside the
disk, and on how or where the plasma attaches to the field lines.
Outstanding issues include whether the disk excludes the stellar 
magnetic field by diamagnetic currents or the field can penetrate and 
thread a large fraction of the disk, whether the threaded 
field remains closed (connecting the star and the disk) or becomes open
by differential shearing, and whether/how magnetically driven wind is
launched from the disk or the magnetosphere/corotation boundary. 

%All authors agree that the disk is truncated at a radius close to $r_m$, and
%that the condition $r_m\simeq r_{\rm co}$ (where $r_{\rm co}$ is the
%corotation radius, at which the Keplerian frequency equals the spin frequency)
%represents an ``equilibrium'' state of the system. 

This paper introduces some new physical effects 
associated with disk accretion onto rotating magnetized stars. 
Most previous papers have, for simplicity, 
assumed that the stellar magnetic dipole is aligned with the spin axis. 
Essentially all previous authors, with the exception of 
Lipunov \& Shakura (1980) (see \S 2.1), have 
assumed that the spin is aligned with the disk angular momentum --- 
There is a good reason for this, as it is natural to think that the 
star attains its spin angular momentum from the accretion disk. 
However, when these assumptions are abandoned, new physical effects are
revealed: We show that under quite general conditions, the stellar magnetic
field can induce disk warping and make the disk 
precess around the spin axis.
As the accreting plasma approaches the magnetosphere,
there is a tendency for the disk plane to be driven away from the 
equatorial plane of the star toward being aligned with the spin axis
(\S 2 and \S 4). 
The origin of the {\it precessional torque} and 
{\it magnetic warping instability}
lies in the interaction between the surface current on the disk
and the horizontal magnetic field (parallel to the disk)
produced by the stellar dipole. The electric current on the
disk is an inevitable consequence of magnetic-field -- disk interaction, 
although its actual form is uncertain and depends on whether the disk is 
diamagnetic or has a vertical field threading it, thus on the dissipative
processes in the disk and the magnetosphere (\S 2). 
The magnetic force on the disk also give rise to vertical and
epicyclic resonances which may drive disk perturbations (\S 3).  
We show that the magnetically driven warping and precession of 
accretion disks can potentially explain a number of outstanding puzzles
related to disk accretion onto magnetized stars (\S 6). 

This paper is organized as follows. 
In \S 2 we calculate the precessional and warping torques on the disk
(or a ring of disk) due to the stellar magnetic field for several 
models based on different assumptions about magnetic 
field -- disk interaction. The torques are different
for different models, but they are always present and are of the same 
order of magnitude. The key physics of the torques are summarized
at the beginning of \S 2. 
Resonances due to magnetic forces on the disk are discussed in \S 3.
In \S 4 we study the dynamics of a warped disk under the influence
of the magnetic torques derived in \S 2.
The criterion for the magnetically driven warping instability (including the
effect of viscosity) is derived, and a magnetic ``Bardeen-Petterson'' effect
is discussed. Section 5 addresses the question of how the spin of the central
star is affected by the warped, precessing disk. 
In \S 6 we discuss/speculate several astrophysical applications of our theory,
including the spin evolution of accretion-powered X-ray pulsars,
quasi-periodic oscillations in low-mass X-ray binaries, 
long-term periodic cycles in several X-ray binaries, and 
the variability and rotation of T Tauri stars. 
We conclude in \S 7 by discussing 
possible future studies along the line initiated in this paper. 

Because of the intrinsic uncertainties associated with the nature
of magnetic field -- disk interaction, in the main text we shall 
focus on generic features and rely on parametrized models. In Appendix
A we discuss a global magnetized disk model. In Appendix B we
consider the issue of calculating the 
magnetosphere radius for nontrivial magnetic field and spin geometry. 
The dynamics of an extreme form of disk (consisting 
of diamagnetic blobs) is examined in Appendix C.

%%%%%%%%%%%%%%%%%%%%%%%%%%%%%%%%%%%%%%%%%%%%%%%%%%%%%%%%%%%%%%%%%%%%%%
\section{MAGNETIC PRECESSIONAL AND WARPING TORQUES}
 
In this section we calculate the magnetic torque on the disk 
(or a ring of disk) surrounding
a rotating magnetic dipole. Because of the uncertainties on how
the magnetic field behaves in the presence of an accretion disk
(see \S 1), we shall consider several different models,
some representing extreme situations.  
Our most ``realistic'' model is  given in \S 2.3.

The basic set-up used in our calculation is shown in Fig.~1.
The disk angular momentum axis $\hatl$ is inclined by an angle $\beta$
with respect to the spin axis $\hato$.
The stellar dipole momentum $\hatmu$ rotates around
the spin axis $\hato$. The angle of obliquity between the magnetic moment 
and the rotation axis is $\theta$. 

The key physics responsible for the magnetic torques on the
disk is as follows:

(i) If the disk is diamagnetic (perfectly conducting) so that
the vertical stellar field (perpendicular to the disk) 
cannot penetrate, an azimuthal screening surface current will
be induced in the disk. This current interacts with the radial magnetic
field from the star, and a vertical magnetic force results. 
While the mean force (averaging over the azimuthal) is zero, 
the uneven distribution of the force induces a net torque 
acting on the ring, making it precess around the spin axis. 
(For a nonrotating star, the disk will precess around the magnetic axis.)
(\S 2.1).

(ii) If there is a vertical field threading the disk (this $B_z$
either comes from the star or is carried intrinsically by the disk and
detached from the star), it will be twisted by the disk rotation
to produce a discontinuous azimuthal field $\Delta B_\phi$ 
across the disk surface, and a radial surface current results. 
The interaction between
this current and the stellar $B_\phi$ (which is not affected by the disk)  
gives rise to a vertical force, and the resulting torque 
tends to misalign the angular momentum of the ring with 
the spin axis (although in some extreme situations, alignment torque
may also result). (For a nonrotating star, the disk normal will be driven 
away from the stellar magnetic axis.) (\S 2.2).

In general, we expect both types of torques to exist on the disk (\S 2.3
and \S 2.4).  

%======================================================================
\subsection{Diamagnetic Disk}

Here we consider the extreme situation where the disk 
is a perfect conductor and has no large-scale 
magnetic field of its own. The inner radius of the disk is located at
$r=r_m$, the magnetospheric radius.
The magnetic field produced by the stellar dipole cannot 
penetrate the disk, and a diamagnetic
surface current is induced. Aly (1980) has found the exact analytic solution
to this model problem (see also Kundt \& Robnik 1980; Riffert 1980).
The magnetic field at a point 
$(r,\phi,z=0)$ (cylindrical coordinates) on the disk surface ($z=0,~r>r_m$)
is given by
\ba
B_r&=&{2\mu\over r^3}\sin\chi\cos(\phi-\phi_\mu)\mp{4\mu\over\pi r^3D}
\cos\chi,\label{br1}\\
B_\phi&=&{\mu\over r^3}\sin\chi\sin(\phi-\phi_\mu),\label{bphi1}\\
B_z&=&0,
\ea
where $\chi$ and $\phi_\mu$ are defined in Fig.~1 (they are both varying 
in time). In eq.~(\ref{br1}), the upper (lower) sign applies to the upper
(lower) disk surface, and the factor $D$ is 
\be
D={\rm max}\left(\sqrt{{r^2/r_m^2}-1},
\sqrt{{2H/r_m}}\right),
\label{Dfunc}\ee
(where $H\ll r_m$ is the half-thickness of the disk). 
The discontinuity in $B_r$ implies a surface current
\be
\bK=-{2c\mu\over \pi^2r^3D}\cos\chi\,\hat\phi
\ee
($\hat\phi$ is the unit vector along the $\phi$ direction; similar notation 
will be used throughout the paper).
Note that this surface current is induced to cancel the $z$-component of the
stellar field, $B_z^{(0)}=-\mu\cos\chi/r^3$; the $r,\phi$-components of the
stellar field [the first terms of eqs.~(\ref{br1}) and (\ref{bphi1})]
do not induce any net surface current. (In fact, they induce
currents on the upper and lower surfaces of the disk, but these currents have
opposite directions. Such currents can lead to the ``squeezing'' of the disk
--- changing the thickness of the disk, but not the ``lifting''.) 
The magnetic force per unit area on the disk
results from the interaction between $\bK$ and $B_r^{(0)}$ (the first
term in eq.~[\ref{br1}]):
\be
\bF={2\mu^2\over\pi^2r^6D}\sin 2\chi\cos(\phi-\phi_\mu)\,\hat z.
\label{force1}\ee
The existence of this vertical magnetic force has already been noted by 
Aly (1980), and 
it is simply the difference in the magnetic pressure,
$B^2/(8\pi)$, between the lower and upper surfaces.
The magnetic torque per unit area is 
\be
\bN=\br\times \bF=-{2\mu^2\over\pi^2r^5D}\sin 2\chi\cos(\phi-\phi_\mu)\hat\phi.
\ee
Clearly, averaging the force over the azimuthal angle, $\langle\bF\rangle_\phi
=(1/2\pi)\int_0^{2\pi}d\phi\, \bF$, gives zero, but the 
azimuthally averaged torque is nonzero, and given by
\be
\langle\bN\rangle_\phi={\mu^2\over\pi^2r^5D}\sin 2\chi(\sin\phi_\mu\hatx
-\cos\phi_\mu\haty)
={2\mu^2\over\pi^2r^5D}\cos\chi\,(\hatmu\times\hatl),
\ee
($\hatmu$ and $\hatl$ are the unit vectors along the dipole moment and
the disk angular momentum, respectively).
For nonrotating star, this result implies that the magnetic torque
tends to make the disk precess around the magnetic axis $\hatmu$. For
a rotating star --- as long as the rotation period is much shorter
than the precession period, we need to average over the spin period.
The following identities will be needed:
\ba
\cos\chi &=&\cos\beta\cos\theta-\sin\beta\sin\theta\sin\omega t,\label{id1}\\
\sin\chi\cos\phi_\mu &=&\sin\theta\cos\omega t,\label{id2}\\
\sin\chi\sin\phi_\mu &=& \sin\beta\cos\theta+\cos\beta\sin\theta\sin\omega t,
\label{id3}\ea
where an appropriate phase for the stellar rotation has been adopted. 
We find that, after averaging over the spin period,
the torque per unit area $\langle\langle\bN\rangle\rangle\equiv
\langle\langle\bN\rangle_\phi\rangle_\omega\equiv
(1/P_s)\int_0^{P_s}dt\,\langle\bN\rangle_\phi$ (where $P_s=2\pi/\omega$
is the spin period) is given by
\be
\langle\langle\bN\rangle\rangle
%={\mu^2\over 2\pi^2r^5D}\sin 2\beta\left(3\cos^2\theta-1\right)\hatx
={\mu^2\over \pi^2r^5D}\cos\beta
\left(3\cos^2\theta-1\right)\,(\hato\times\hatl).
\label{torque1}\ee
The magnetic torque on a ring of radius $r$ and width $dr$ is simply 
$d\bT=2\pi r\langle\langle\bN\rangle\rangle dr$.
The angular momentum of the ring is $(2\pi r\Sigma dr)(r^2\Omega)$ (where
$\Sigma$ is the surface density of the disk, and $\Omega$ is the
orbital angular frequency). Thus the effect of the magnetic torque is to 
make the ring precess around the spin axis $\hato$ at an angular frequency
\be
{\bf \Omega}_{\rm prec}={\mu^2\over \pi^2r^7\Omega\Sigma D}\cos\beta\,
\left(3\cos^2\theta-1\right)\,\hato.
\ee
Note that the sign of $\cos\beta \left(3\cos^2\theta-1\right)$ determines
whether ${\bf \Omega}_{\rm prec}$ is along $\hato$ or opposite to it.
This result has been obtained previously by Lipunov \& Shakura (1980)
\footnote{Lipunov \& Shakura (1980) also argued that 
for $\cos^2\!\theta<1/3$, the minimum of the 
interaction energy between the central
dipole and the field generated by the disk current is achieved at
$\beta=0^\circ$, while for $\cos^2\!\theta>1/3$, the minimum energy corresponds
to $\beta=90^\circ$. They therefore suggested that in the latter case
($\cos^2\!\theta>1/3$), the disk tends to evolve into the $\beta=90^\circ$
state. However, if eq.~(\ref{torque1}) is the only torque present
in the disk, it is not clear how $\beta$ can change. Moreover, the 
``magnetic Bardeen-Petterson'' effect always tends to align $\hatl$ 
and $\hato$, independent of the sign of ${\Omega}_{\rm prec}$ 
(see \S 4.2). I became aware of the papers by
Lipunov et al.~(see also Lipunov, Sem\"enov \& Shakura 1981)
in mid-February 1999, at which point 
this paper was nearly finished. For completeness and pedagogical reason, 
I have decided to keep this subsection (\S 2.1) in its original form. 
I thank Dr.~Brad Hansen (CITA) for drawing my attention to the papers by 
Lipunov et al.}.

%=============================================================
\subsection{Magnetically Threaded Disk}

We now consider the opposite limit in which the stellar magnetic field
rapidly penetrates the disk (on a timescale shorter than 
the dynamical time of the disk) (see \S 2.3). Because of the shear between
the disk and the plasma outside the disk, the threaded vertical 
field is winded to produce an azimuthal field which has different signs
on the upper and lower surfaces of the disk. We thus adopt the
following {\it ansatz} for the magnetic field in the disk\footnote{
In principle, an additional radial field $\Delta B_r$ could be generated
(with opposite signs in the upper and lower disk surfaces) 
when the radial inflow of the disk drags the threaded $B_z$. 
This discontinuous $\Delta B_r$ will give rise to 
a precessional torque similar to that derived in \S 2.1. 
Since we expect $\Delta B_r$ to be proportional to the radial velocity, 
it is neglected in eq.~(\ref{br2}).}:
\ba
B_r&=&{2\mu\over r^3}\sin\chi\cos(\phi-\phi_\mu),\label{br2}\\
B_\phi&=&{\mu\over r^3}\sin\chi\sin(\phi-\phi_\mu)\pm\zeta{\mu\over
r^3}\cos\chi,\label{bphi2}\\
B_z&=&-{\mu\over r^3}\cos\chi.\label{bz2}
\ea
In (\ref{bphi2}), the second term represents the field produced by
the twisting of $B_z$, and the upper (lower) sign corresponds to the value
at the upper (lower) disk surface\footnote{
There could be an additional contribution to
$B_\phi$ on the right-hand side of (\ref{bphi2}) due to the 
shearing of the stellar $B_r$ by the differential rotation. This 
contribution is not included for two reasons. First, the stellar $B_r$ may 
not be able to penetrate the thin disk; Second, even if a component of $B_r$ 
(e.g., the static component; see \S 2.3) penetrates the disk, the induced 
$\Delta B_\phi$ is expected to be smaller than that due to the threaded 
$B_z$. To see this, we write, schematically, 
$${\partial \Delta B_\phi\over
\partial t}=r\left(B_z{\partial\Omega\over\partial z}+B_r
{\partial\Omega\over\partial r}\right)-{\Delta B_\phi\over\tau_{\rm diss}},$$
where $B_z$ and $B_r$ are the threaded fields, $\tau_{\rm diss}$ is the
effective dissipation time. In steady state, we find
$$\Delta B_\phi=\mp \zeta B_z-{3\over 2}
\tau_{\rm diss}\Omega B_r,$$
where $\zeta=(r/H_B)\tau_{\rm diss}
(\Omega-\omega)$ (for closed field configurations) or
$\zeta=(r/H_B)\tau_{\rm diss}\Omega$ (for open field configurations)
($H_B$ is the vertical scale in which $\Omega$ varies). 
Note that the shearing of $B_r$ does not produce any surface current.
For $H_B\ll r$, we can drop the term proportional $B_r$ in $\Delta B_\phi$,
and obtain $\Delta B_\phi=\mp \zeta B_z$. 
}.
The quantity $\zeta$ specifies the
azimuthal pitch of the field line. In general we expect 
$|\zeta|\lo 1$ (e.g., Sturrock \& Barnes 1972; Lovelace et al.~1995 and
references therein), but its actual value or form depends on details
of the dissipative processes involved in the disk-magnetic field
interactions. If the stellar magnetic field threads
the disk in a closed configuration (e.g., Ghosh \& Lamb 1979a,b; Wang
1987,1995), we expect $\zeta\propto (\Omega-\omega)$ so that $\zeta>0$ for
$\Omega>\omega$ and $\zeta<0$ for $\Omega<\omega$. But it has been
argued that the differential shearing and the plasma flowing from the disk 
into the overlying magnetosphere will blow the field lines open and 
maintain them in a open configuration (e.g., 
%Aly 1985,1991; Sturrock 1991
Arons 1987; Newman et al.~1992; Lynden-Bell \& Boily 1994;
Lovelace et al.~1995 and references therein), 
in which case we expect $\zeta$ to be positive and of order unity. 
For our purpose in this paper, $\zeta$ is simply a
dimensionless number or function\footnote{Even 
in models (e.g., Shu et al.~1994) 
where the disk is largely diamagnetic with no intrinsic $B_z$,
the stellar field must penetrate the boundary layer. In this case, one would
imagine that $\zeta$ is nonzero only near the magnetosphere boundary.}.
Also note that because of 
the screening currents in the disk (such as those discussed in \S 2.1)
or in the magnetosphere boundary, the threaded field 
may be smaller than the vacuum field produced by the dipole. 
But since our result (see below) depends on $\zeta\mu^2$, we can easily 
absorb the screening effect into the definition of $\zeta$.

The surface current on the disk corresponding to 
the field (\ref{br2})-(\ref{bz2}) is
\be
\bK=-{\zeta c\mu\over 2\pi r^3}\cos\chi\,\hatr.
\ee
The interaction of $\bK$ with $B_z$ gives rise to an azimuthal force, 
which tends to slow down (or speed up if $\omega>\Omega$ and the field lines
are in a closed configuration) the fluid motion, 
transferring angular momentum between the disk and the overlying 
magnetosphere or the star. This is the familiar magnetic breaking torque which
has been included in previous studies of aligned magnetized disk 
(e.g., Ghosh \& Lamb 1979a,b; Lovelace et al.~1995; Wang 1995; Lai 1998).
Here we are interested in the 
vertical magnetic force $F_z$ (per unit area), 
resulting from the interaction between
$\bK$ and the azimuthal field produced by the stellar dipole 
(the first term in [\ref{bphi2}]):
\be 
F_z=-{\zeta\mu^2\over 4\pi r^6}\sin 2\chi\sin(\phi-\phi_\mu).
\label{force2}\ee
This is simply the difference in $B^2/(8\pi)$ below and above the disk. 
The magnetic torque associated with $F_z$ is given by
\be
\bN={\zeta\mu^2\over 4\pi r^5}\sin 2\chi
\sin (\phi-\phi_\mu)\,\hat\phi.
\ee
Averaging over $\phi$, we have
\be
\langle\bN\rangle_\phi=-{\zeta\mu^2\over 8\pi r^5}\sin 2\chi
(\cos\phi_\mu\hatx+\sin\phi_\mu\haty)
=-{\zeta\mu^2\over 4\pi r^5}(\hatl\cdot\hatmu)\left[\hatmu-(\hatmu\cdot
\hatl)\hatl\right].
\label{torquephi1}\ee
For a nonrotating star, this torque tends to pull the disk normal 
vector $\hatl$ toward being perpendicualr to
the magnetic axis $\hatmu$ (assuming $\zeta>0$),
thus making the disk plane parallel to the stellar field lines.
For a rotating star, averaging over the spin period and using
the indentities (\ref{id1})-(\ref{id3}), we find
\ba
\langle\langle\bN\rangle\rangle &=&
-{\zeta\mu^2\over 16\pi r^5}\sin 2\beta \left(3\cos^2\theta-1\right)\,\haty
\nonumber\\
&=&-{\zeta\mu^2\over 8\pi r^5}\cos\beta \left(3\cos^2\theta-1\right)
\left[\hato-(\hato\cdot\hatl)\hatl\right].
\label{torque2}\ea
In the absence of other forces, this magnetic torque will change 
the tilt angle $\beta$ of the ring (at radius $r$) according to
\be
{d\beta\over dt}={\zeta\mu^2\over 16\pi r^7\Omega\Sigma}\sin 2\beta
\left(3\cos^2\theta-1\right).
\label{dbeta1}\ee
Thus, depending on the sign of $\zeta\cos\beta (3\cos^2\theta-1)$,
the angle $\beta$ can increase or decrease: For 
$\zeta (3\cos^2\theta-1)>0$, the torque drives the disk toward the
perpendicular configuration ($\beta=90^\circ$), while for 
$\zeta (3\cos^2\theta-1)<0$, it drives the disk toward alignment
($\beta=0^\circ$) or anti-alignment ($\beta=180^\circ$).
In particular, when $\zeta (3\cos^2\theta-1)>0$, the aligned
configuration is unstable against the growth of disk tilt angle
--- This is the {\it magnetic warping instability} (see \S 2.3 and \S 2.4). 

It is instructive to understand the difference between the 
situation studied here and that of a diamagnetic disk considered in \S2.1.
In both cases, a vertical magnetic force is exerted on the disk.
Although eqs.~(\ref{force1}) and (\ref{force2})
appear similar, they have very different spin-averaged behavior. 
For a diamagnetic disk, we have
\be
\langle F_z\rangle_{\omega}={\mu^2\over\pi^2 r^6D}\sin
2\beta\left(3\cos^2\theta-1\right)\,\sin\phi.
\ee
Thus the force on the $y>0$ side of the disk has a different sign 
from that on the $y<0$ side, giving rise to a torque which is 
along the $x$-axis (perpendicular to both $\hatl$ and $\hato$).
For a magnetically threaded disk, we have
\be
\langle F_z\rangle_{\omega}={\zeta\mu^2\over 8\pi r^6}\sin
2\beta\left(3\cos^2\theta-1\right)\,\cos\phi.
\ee
The force has different signs for $x>0$ and $x<0$, and the torque
is along the $y$-axis (in the same plane as $\hatl$ and $\hato$). 

%==========================================================
\subsection{A Hybrid ``Realistic'' Model}

The results of \S 2.1 and \S 2.2 represent two opposite,
extreme situations, and are not likely to be realistic. 
The magnetic field configuration in a purely diamagnetic 
disk (see \S 2.1) is prone to Kelvin-Helmholtz instability and 
reconnection. 
%(e.g., Aly 1980). 
It has been argued that the non-linear
development of the these processes could lead to partial threading 
of the magnetic field through the disk (e.g., Ghosh \& Lamb 1979a,b; 
Wang 1987,~1995), although the extent of the field-threading region is 
uncertain and the closed configuration advocated by Ghosh and Lamb is probably
unrealistic (e.g., Arons 1987; Shu et al.~1994;
Lovelace et al.~1995). 
In any event, near the magnetosphere boundary, the vertical 
stellar field must certainly thread the disk plasma.
On the other hand, the magnetically thread disk model considered in \S 2.2
requires that the stellar field penetrates the disk almost instantaneously 
(compared to the spin period and the orbital period). This is 
unrealistic. The timescale for field penetration is uncertain, but it 
cannot be shorter than the dynamical time of the disk; it may be 
as long as the disk thermal time (E.~T. Vishniac 1999, private communication; 
see Park \& Vishniac 1996 and references therein).

The vertical vacuum field produced by the stellar dipole 
on the disk can be written as the sum of a static component and a 
time-varying component:
\be
B_z^{(0)}=-{\mu\over r^3}\cos\chi=-{\mu\over r^3}\cos\beta\cos\theta
+{\mu\over r^3}\sin\beta\sin\theta\sin\omega t.
\label{bz0}\ee
While it is possible that the static field can penetrate the disk 
by allowing sufficient time for Kelvin-Helmholtz instability/reconnection 
to grow, it is almost certain that the variable field will be shielded 
out of the disk by the screening current.
We therefore consider the following hybrid model, which we consider
to be more realistic than the situations studied in \S 2.1 and \S 2.2:
The static component of the vertical stellar field threads the disk, 
while the time-varying component is screened out by the disk.
% which, however, responds diamagnetically to the 
%varying component of the vertical field. 
The winding of the threaded field will produce 
an azimuthal field and a radial surface current, while the
variable vertical field will induce a shielding azimuthal surface current 
and discontinuous radial field --- this radial field can be obtained 
by appropriately modifying Aly's solution as given in \S 2.  
The magnetic field on the disk is then given by
\footnote{In eq.~(\ref{br3}) we have neglected a possible component
of $B_r$ generated by radial infall of the threaded $B_z$; See footnote 2.
Also, in (\ref{bphi3}) we have ignored a possible toroidal field 
generated by the shearing of the stellar $B_r$; see footnote 3.}
\ba
B_r&=&{2\mu\over r^3}\sin\chi\cos(\phi-\phi_\mu)\pm{4\mu\over\pi
r^3D}\sin\beta\sin\theta\sin\omega t,\label{br3}\\
B_\phi&=&{\mu\over r^3}\sin\chi\sin(\phi-\phi_\mu)\pm\zeta{\mu\over
r^3}\cos\beta\cos\theta,\label{bphi3}\\
B_z&=&-{\mu\over r^3}\cos\beta\cos\theta.\label{bz3}
\ea
In (\ref{br3}), $B_r$ is the sum of the vacuum dipole field and the field
produced by the azimuthal screening current (which is induced to cancel 
the variable part of $B_z^{(0)}$); in (\ref{bphi3}),
$B_\phi$ is the sum of the vacuum dipole field and 
the field induced by the winding of the constant $B_z$ (see \S 2.2 for
the property of $\zeta$). The surface current on the disk is
\be
\bK={2c\mu\over\pi^2r^3D}\sin\beta\sin\theta\sin\omega t\,\hat\phi-
{\zeta c\mu\over 2\pi r^3}\cos\beta\cos\theta\,\,\hat r.
\ee
The vertical magnetic force (per unit area) on the disk is then given by
\be
F_z=-{4\mu^2\over\pi^2 r^6D}\sin\beta\sin\theta
\sin\omega t\sin\chi\cos(\phi-\phi_\mu)
-{\zeta\mu^2\over 2\pi r^6}\cos\beta\cos\theta\sin\chi\sin(\phi-\phi_\mu).
\label{force3}\ee
The corresponding torque (per unit area) acting on the disk, averaged over 
the ring, is given by
\be
\langle\bN\rangle_\phi=-{2\mu^2\over\pi^2 r^5D}\sin\beta\sin\theta
\sin\omega t\,(\hatmu\times\hatl)
-{\zeta\mu^2\over 4\pi r^5}\cos\beta\cos\theta\,\left[\hatmu
-(\hatmu\cdot\hatl)\hatl\right].
\label{torquephi}\ee
For nonrotating star, eq.~(\ref{torquephi}) reduces to eq.~(\ref{torquephi1})
(note that $\cos\chi=\cos\beta\cos\theta$ when $\omega=0$):
The torque tends to make the disk plane align with the stellar field lines
(i.e., $\hatl$ perpendicular to $\hatmu$). 
Averaged over the stellar rotation, the torque can be written as
\be
\langle\langle\bN\rangle\rangle=
\langle\langle\bN\rangle\rangle_{\rm prec}+
\langle\langle\bN\rangle\rangle_{\rm warp},
\label{torque3a}\ee
where the precessional torque and the warping torque are given by
\ba
\langle\langle\bN\rangle\rangle_{\rm prec}
&=&-{\mu^2\over\pi^2
r^5D}\cos\beta\sin^2\!\theta\,(\hato\times\hatl),\label{torque3b}\\
\langle\langle\bN\rangle\rangle_{\rm warp}
&=&-{\zeta\mu^2\over 8\pi r^5}\sin 2\beta\cos^2\!\theta\,\haty
=-{\zeta\mu^2\over 4\pi r^5}\cos\beta\cos^2\!\theta\,\left[
\hato-(\hato\cdot\hatl)\hatl\right].
\label{torque3c}\ea
Clearly, the diamagnetic feature of the disk induces a precessional
torque (eq.~[\ref{torque3b}]), with the precession angular frequency given by
\be
{\bf\Omega}_{\rm prec}=-{\mu^2\over\pi^2
r^7\Omega\Sigma D}\cos\beta\sin^2\!\theta\,\hato.
\label{prec3}\ee
On the other hand, the disk-threading field
gives rise to a torque (eq.~[\ref{torque3c}]) which 
changes $\beta$ at a rate given by
\be
{d\beta\over dt}={\zeta\mu^2\over 8\pi r^7\Omega\Sigma}
\sin 2\beta\cos^2\!\theta.
\label{warp3}\ee
Note that the disk precession is along the direction 
$(-\cos\beta\,\hato)$, i.e., the precession 
is always in the opposite sense as the orbital motion of disk.
Equation (\ref{warp3}) reveals the {\it magnetic warping 
instability}: the spin-orbit inclination angle 
$\beta$ increases when $\zeta\cos\beta>0$ and decreases when
$\zeta\cos\beta>0$. Thus, for $\zeta>0$ (the most likely case; see \S 2.2), 
the warping torque always tends to drive the disk toward the 
configuration where $\hatl$ is perpendicular to $\hato$.

%==========================================================
\subsection{General Consideration}

The preceding subsections assume that the vertical field 
that threads the disk originates from the star. But it is 
important to note that this is not a requirement for 
the existence of the warping torque. Indeed, models of hydromagnetic
driven outflows from disks 
%in a variety of systems (protostars, neutron stars and black holes) 
are predicated on the existence of large-scale poloidal 
magnetic field that threads the disk --- this field could have been
advected inward by the accreting flow or generated in the disk (see, e.g.,
Blandford 1989 for a review).

%(although this tends to produce odd fields,
%with opposite signs on both sides of the disk). 

Suppose there is vertical $B_z$ (assumed to be time-independent)
which threads the disk. The warping torque results from the interaction of 
the stellar $B_\phi$ and the radial surface current on the disk 
induced by the twisting of the threaded $B_z$. 
%Whenever there is a vertical field, $B_z$, threading the disk --- the field
%could come from the star or is simply carried by the disk itself, differential
%shearing will inevitably produce a discontinuous azimuthal field and 
%a radial current. 
Thus in general, we can write the azimuthal field on the disk as 
\be
B_\phi={\mu\over r^3}\sin\chi\sin(\phi-\phi_\mu)\mp\zeta B_z,
\ee
with $\zeta$ a positive dimensionless number of order or less than unity.
The vertical force on the disk is given by
\be
F_z={\zeta\mu B_z\over 2\pi r^3}\sin\chi\sin(\phi-\phi_\mu).
\label{force4}\ee
The averaged (warping) torque is then
\be
\langle\langle\bN\rangle\rangle={\zeta\mu B_z\over 4\pi r^2}\sin\beta
\cos\theta\,\haty.
\label{torque4}\ee
The model considered in \S 2.3 corresponds to 
$B_z=-\mu\cos\beta\cos\theta/r^3$.

The precessional torque results from the interaction of the ring of 
diamagnetic screening current and the radial magnetic field produced 
by the star. Assume that the surface current has the form
\be
\bK=\left(K_{\phi 1}+K_{\phi 2}\sin\omega t\right)\,\hat\phi.
\ee
(In general, both the static current $K_{\phi 1}$ and 
the time-dependent current $K_{\phi 2}\sin\omega t$ are possible as
the disk responds to the external field that tries to enters 
the disk.) The vertical magnetic force is 
\be
F_z=-{2\mu\over cr^3}\left(K_{\phi 1}+K_{\phi 2}\sin\omega t\right)\sin\chi
\cos(\phi-\phi_\mu),
\label{force5}\ee
and the resulting averaged (precessional) torque is
\be
\langle\langle\bN\rangle\rangle=-{\mu\over cr^2}\left(K_{\phi 1}\sin\beta
\cos\theta+{K_{\phi 2}\over 2}\cos\beta\sin\theta\right)\,\hatx.
\label{torque7}\ee
The model considered in \S 2.3 corresponds to $K_{\phi 1}=0$ and
$K_{\phi 2}=(2c\mu/\pi^2 r^3D)\sin\beta\sin\theta$, while in the model 
considered in \S 2.1, we have $K_{\phi 1}=-(2c\mu/\pi^2r^3D)
\cos\beta\cos\theta$ and $K_{\phi 2}=(2c\mu/\pi^2 r^3D)\sin\beta\sin\theta$.

Finally, we note that in our calculations of the warping torques
(\S 2.2-\S2.4) we have assumed that the azimuthal pitch $\zeta$ 
is stationary in time. An alternative scenario has been proposed 
(Aly \& Kuijpers 1990; van Ballegooijen 1994),
where the threaded field lines are being constantly wound up, with 
occasional field reconnection which releases the stored-up magnetic 
energy. Even for such a time-dependent magnetic field structure
(apart form the time-dependence of the rotating dipole field), 
it is likely that magnetic torques (similar to those calculated in 
this paper) will exist, although this possible complication will be ignored
in this paper.

\subsection{Disk Consisting of Diamagnetic Blobs}

The previous subsections treat the disk as a continuum. It   
has been suggested that under certain conditions, 
the accretion disk may be lumpy, consisting of diamagnetic blobs 
(Vietri \& Stella 1998; see King 1993). Such inhomogeneous accretion 
flow may arise from the nonlinear development of various plasma 
instabilities (e.g., Kelvin-Helmholtz instability;
Arons \& Lea 1980) associated with the disk and the magnetosphere.
Vietri \& Stella (1998) studied the motion of
diamagnetic blobs orbiting a central star with an inclined magnetic dipole
and showed that magnetic drag force (Drell, Foley \& Ruderman 1965)
on the blob can induce vertical resonances near the 
corotation radius of the disk.

We have calculated the magnetic torque on the accretion disk
consisting of individual diamagnetic blobs. 
It can be shown that independent of the vertical resonances, there exists a
magnetic torque which tends to induce tilt on the orbit of the blob. 
Since there are considerable uncertainties associated with such 
an extreme form of accretion disk, we relegate the calculation to 
Appendix C.

%%%%%%%%%%%%%%%%%%%%%%%%%%%%%%%%%%%%%%%%%%%%%%%%%%%%%%%%%%%%%%%%%%
\section{MAGNETICALLY DRIVEN RESONANCES}

The discussion in \S 2 has neglected possible resonances in the interaction
between the disk the rotating central dipole. Here 
we consider these resonances.

\subsection{Vertical and Epicyclic Resonances}

The bending mode of the disk is characterized by a small vertical 
displacement $Z(r,\phi,t)$, with the equation of motion
\be
{d^2Z\over dt^2}=
\left({\partial\over\partial t}+\Omega{\partial\over\partial\phi}\right)^2Z
=-\Omega_z^2 Z+{1\over\Sigma}F_z,
\label{eom1}\ee
where $\Omega$ is the orbital angular frequency, 
the term $-\Omega_z^2z$ represents vertical the gravitational restoring 
force (for a Keplerian disk, $\Omega_z=\Omega$), 
and $F_z$ is the magnetic force (per unit area) as calculated in
\S 2. Pressure, viscous force and self-gravity have been neglected.  

Consider first the magnetic force given by eq.~(\ref{force4}), and write it as
\be 
F_z={\zeta\mu B_z\over 2\pi r^3}\Bigl[\sin\theta\cos\omega t\sin\phi
-\left(\sin\beta\cos\theta+\cos\beta\sin\theta\sin\omega t\right)
\cos\phi\Bigr].
\label{force6}\ee
There exists the following possible vertical resonances:
\ba
&&\Omega_z=\Omega\quad\qquad\qquad~~~~~
({\rm for }~~\sin\beta\cos\theta\neq 0)\label{res1}\\
&&\omega-\Omega=\pm\Omega_z\qquad\qquad({\rm for }~~\sin\theta\neq 0~~
{\rm and}~~\beta\neq \pi)\label{res2}\\
&&\omega+\Omega=\Omega_z\qquad\qquad~~
({\rm for }~~\sin\theta\neq 0~~{\rm and}
~~\beta\neq 0).\label{res3}
\ea
These resonances can be easily understood: (1) 
When $\sin\beta\cos\theta\neq 0$,
there exists a static distribution ($\propto \cos\phi$) of vertical force
field, and a fluid element feels the force $\propto\cos\Omega t$; this 
explains the $\Omega=\Omega_z$ resonance. (2) When $\sin\theta\neq 0$, 
there exists a time-dependent vertical force field ($\propto\sin\omega t$ 
or $\cos\omega t$). A fluid element, traveling with angular velocity $\Omega$,
would experience vertical forces with frequency $(\omega\pm\Omega)$
[for $\beta=0$, only $(\omega-\Omega)$ is possible, while for
$\beta=180^\circ$, only $(\omega+\Omega)$ is possible]; this explains
the $\omega\pm\Omega=\pm\Omega_z$ resonances. 

Now consider the magnetic force given by eq.~(\ref{force5}). 
The force associated with $K_{\phi 1}$ has the form 
$\sin\phi$ and $\cos(\phi\pm\omega t)$, which would give rise
to the $\Omega_z=\Omega$ resonance and the 
$\omega\pm\Omega=\pm\Omega_z$ resonances as in eqs.~(\ref{res1})-(\ref{res3}).
The force associated with $K_{\phi 2}$ has a time-dependence of the form
\ba
\sin\omega t\sin\chi\cos(\phi-\phi_\mu)&=&
{1\over 2}\cos\beta\sin\theta\sin\phi+
\sin\beta\cos\theta\sin\omega t\sin\phi\nonumber\\
&&+{1\over 2}\sin\theta\Bigl(\sin 2\omega t\cos\phi-\cos\beta\cos 2\omega t
\sin\phi\Bigr).
\ea
This gives rise to the following resonances:
\ba
&&\Omega_z=\Omega\quad\qquad\qquad~~~~~~~\,
({\rm for }~~\cos\beta\sin\theta\neq 0)\label{reso1}\\
&&\omega\pm \Omega=\pm\Omega_z\qquad\qquad~~\,({\rm for }~~\sin\beta\cos\theta
\neq 0)\label{reso2}\\
&&2\omega-\Omega=\pm\Omega_z\qquad\qquad~
({\rm for }~~\sin\theta\neq 0~~{\rm and}~~\beta\neq \pi)\label{reso3}\\
&&2\omega+\Omega=\Omega_z\qquad\qquad~~~
({\rm for }~~\sin\theta\neq 0~~{\rm and}~~\beta\neq 0).\label{reso4}
\ea
The new resonances, $2\omega\pm\Omega=\pm\Omega_z$, come about 
because the magnetic field varies as $\cos\omega t$ or $\sin\omega t$, and
the screening current also varies as $\cos\omega t$ or $\sin\omega t$.

%Therefore resonance can also occur when $\Omega\pm 2\omega=\pm\Omega_z$,
%i.e., \be \Omega+\Omega_z=2\omega,\ee
%or $\Omega=\omega$ for Keplerian orbit.
%At both the $\Omega=\omega$ and $2\Omega=\omega$ resonances, 
%in addition to eq.~(\ref{torque7}), there are ``resonant'' torques 
%which depend on $\phi_0$ (analogous to eq.~[\ref{torque5}]).

In addition to the vertical resonances discussed above, 
epicyclic resonance can also arise from the magnetic 
field -- disk interaction. Consider the field given by 
eqs.~(\ref{br3})-(\ref{bz3}). The interaction between $K_\phi$
with $B_z$ gives rise to a radial force 
\be
F_r=-{\mu^2\over 2\pi^2r^6D}\sin 2\beta\sin 2\theta\sin\omega t.
\ee
Clearly, epicyclic resonance occurs when
\be
\kappa=\omega\qquad\qquad({\rm for}~~\sin 2\beta\sin 2\theta\neq 0),
\label{epy}\ee
where $\kappa$ is the epicyclic frequency. For a Keplerian disk, 
$\kappa=\Omega$, (\ref{epy}) is a corotation resonance.

We note that this paper deals only with dipole field from the star. When 
higher-order multipole fields are considered, it is conceivable that 
additional resonances can arise.

\subsection{Magnetic Torques at Resonances}

What are the consequences of the magnetically driven resonances? 
We have not investigated this issue in detail. The resonances
may act as an extra source (in addition to the non-resonant
warping torques discussed in \S 2 and \S 4) for generating bending waves and
spiral waves in the disk. Near the resonances, fluid elements 
undergo large out-of-plane and radial excursions, which may lead to
thickening of the disk\footnote{This may be analogous to the 
Lorentz resonances (which occur when charge particles
move around a rotating magnetic field) in the jovian ring
(Burns et al.~1985; Schaffer \& Burns 1992). 
However, because of the fluid nature of the disk, 
the resonances may not lead to sharp edges in the disk.}.

We can get some insight into the resonances by calculating the magnetic
torques at the resonant radii. Consider first the torque associated with 
the force in eq.~(\ref{force6}). For a Keplerian disk, 
the condition $\Omega_z=\Omega$ is always satisfied.
%It is not clear whether there is any significance to this, other than 
%the fact that 
When $\sin\beta\cos\theta\neq 0$,
there always exists a net (averaged) torque acting on the fluid element, 
pulling its orbital angular momentum axis toward (or away from) the spin 
axis (eq.~[\ref{torque4}]). The condition $\Omega+\Omega_z=\omega$
(or $2\Omega=\omega$ for a Keplerian disk) specifies a special radius
in the disk. To calculate the net torque on the fluid element
at this resonant radius, we cannot average $\phi$ and the spin period
independently (as done in \S 2). Instead 
we write the torque $\bN=\br\times\bF$ as:
\ba
\bN&=&{\zeta\mu B_z\over 4\pi r^2}\biggl\{
\Bigl[\sin\theta\cos\omega t (1-\cos 2\phi)-(\sin\beta\cos\theta+\cos\beta
\sin\theta\sin\omega t)\sin 2\phi\Bigr]\,\hatx\nonumber\\
&&+\Bigl[-\sin\theta\cos\omega t\sin 2\phi+(\sin\beta\cos\theta
+\cos\beta\sin\theta\sin\omega t)(1+\cos 2\phi)\Bigr]\,\haty\biggr\}.
\ea
Following a fluid element we have $\phi=\Omega t+\phi_0$. 
Averaging over time, we find that, at the $2\Omega=\omega$ resonance, 
the torque is given by
\be
\langle\langle\bN\rangle\rangle={\zeta\mu B_z\over 4\pi r^2}\sin\beta
\cos\theta\,\haty
-{\zeta\mu B_z\over 8\pi r^2}(1+\cos\beta)\sin\theta
\,(\hatx\cos 2\phi_0+\haty\sin 2\phi_0).
\label{torque5}\ee
Thus for each fluid element on the resonant radius, the averaged
torque is modified from the ``nonresonant'' value (the first term
in eq.~[\ref{torque5}]). 
%This clarifies the significance of the $2\Omega=\omega$ resonance. 
Since different fluid elements in the same ring have different 
values of $\phi_0$, they will experience different torques; if they are allowed
to move independent of each other, the ring will eventually disperse. However,
if there is strong coupling between the different elements on the ring so that
the ring evolve dynamically as an identity, then one should average over 
$\phi_0$, and eq.~(\ref{torque5}) reduces to (\ref{torque4}).
We can similarly consider the torque associated with 
the magnetic force given in eq.~(\ref{force5}). 
At both the $\Omega=\omega$ and $2\Omega=\omega$ resonances, 
in addition to eq.~(\ref{torque7}), there are ``resonant'' torques 
which depend on $\phi_0$ (analogous to eq.~[\ref{torque5}]).

We note that the ``nonresonant'' torque is nonzero only when 
$\sin\beta\neq 0$ (misaligned spin-orbit), 
while at the $2\Omega=\omega$ or $\Omega=\omega$
resonances, the torque can be nonzero even when $\beta=0$. 
Also note that in general, the ``resonant'' torque is of the same order of
magnitude as the ``nonresonant'' torque. We shall not consider these
resonances in the rest of the paper.

%%%%%%%%%%%%%%%%%%%%%%%%%%%%%%%%%%%%%%%%%%%%%%%%%%%%%%%%%%%%%%%%%%%%%%
\section{DYNAMICS OF WARPING AND PRECESSING DISK}

We now study the dynamics of the disk under the influence
of the magnetic torques calculated in \S 2.
For concreteness, we shall use the torque expressions
of \S 2.3; using other expressions of \S 2 would give similar results
(although the dependence on angles would be different).
We are particularly interested in whether the warping instability 
can operate in the presence of disk viscosity.

%=================================================================
\subsection{Criterion for the Warping Instability}

Our starting point is the evolution equation for disk tilt $\hatl (r,t)$ 
(the unit vector perpendicular to the disk annulus at radius $r$) given by
Pringle (1992) (see also Papaloizou \& Pringle 1983)
\footnote{Recent study (Ogilvie 1999) indicates that in the nonlinear regime, 
the warping equation needs to be modified. In effect, $\nu_1$ and $\nu_2$ 
are not the usual vertically averaged viscosities and may depend on the 
amplitude of the warp.}:
\ba
&&{\partial\hatl\over\partial t}+\left[V_r-{\nu_1\Omega'\over\Omega}
-{1\over 2}\nu_2{(\Sigma r^3\Omega)'\over\Sigma r^3\Omega}\right]
{\partial\hatl\over\partial r}\nonumber\\
&&\quad\qquad ={\partial\over\partial r}\left({1\over 2}\nu_2{\partial\hatl
\over\partial r}\right)+{1\over 2}\nu_2\left|{\partial\hatl\over\partial r}
\right|^2+{{\bf N}\over\Sigma r^2\Omega}.
\label{warp}\ea
Here $V_r$ is the radial velocity of the flow, $\Omega'\equiv d\Omega/dr$,
$\nu_1$ is the usual disk viscosity (measuring the $r$-$\phi$ stress),
and $\nu_2$ is the viscosity (measuring the $r$-$z$ stress)
associated with reducing disk tilt.
We assume that the timescale
for $\hatl$ to change is much longer than the spin period, so that 
$\bN$ is the averaged torque (per unit area) as calculated in \S 2.3
(The notation $\langle\langle\cdots\rangle\rangle$ has been suppressed). 
Using the following relations for a Keplerian disk:
\be
V_r=-{3\nu_1\over 2r}{\cal J}^{-1},~~~~
\Sigma={\dot M\over 3\pi\nu_1}{\cal J},
\label{vr}\ee
where ${\cal J}$ is a function of $r$ which approaches unity in the region 
far from the inner edge of the disk (see Appendix A), and assuming
that $\nu_2/\nu_1$ is independent of $r$, we reduce eq.~(\ref{warp}) to
\ba
&&{\partial\hatl\over\partial t}-\left[{3\nu_2\over 4r}\left(1+
{2r\cJ'\over 3\cJ}\right)+{3\nu_1\over 2r}\left(\cJ^{-1}-1\right)\right]
{\partial\hatl\over\partial r}\nonumber\\
&&\quad\qquad ={1\over 2}\nu_2{\partial^2\hatl\over\partial r^2}
+{1\over 2}\nu_2\left|{\partial\hatl\over\partial r}
\right|^2+{{\bf N}\over\Sigma r^2\Omega}.
\label{warp1}\ea
We rewrite the magnetic torques in eqs.~(\ref{torque3b}), (\ref{torque3c}) as
\ba
{\bN_{\rm prec}\over\Sigma r^2\Omega}&=&-\Omega_p\cos\beta\,\,\hato\times\hatl,
\,~~~~~~~~~\Omega_p={\mu^2\over\pi^2r^7\Omega\Sigma D}\sin^2\theta,
\label{omegap}
\\
{\bN_{\rm warp}\over\Sigma r^2\Omega}&=&-\Gamma_w\cos\beta\sin\beta\,\haty,
~~~~~~~~\Gamma_w={\zeta\mu^2\over 4\pi
r^7\Omega\Sigma}\cos^2\theta.\label{gammaw}
\ea
At this stage $\hatl$ is no longer useful as a coordinate axis, 
so we consider a different cartesian coordinate system where 
$\hato$ is the $z'$-axis (see Fig.~1). In this coordinate system, we write
\be 
\hatl=(\sin\beta\cos\gamma,\sin\beta\sin\gamma,\cos\beta),
\ee
which defines the local twist angle $\gamma$. For small tilt angle 
($\beta\ll 1$), eq.~(\ref{warp1}) simplifies to
\be
{\partial W\over\partial t}-{3\nu_2\over 4r}{\partial W\over\partial r}
={1\over 2}\nu_2{\partial^2W\over\partial r^2}
-i\Omega_pW+\Gamma_wW,
\label{warpw}\ee
where $W\equiv \beta\, {\rm e}^{i\gamma}$, and we have set $\cJ=1$ (far from
the inner edge of the disk) -- using a more rigorous $\cJ$ would
only affect the numerical coefficient in front of  the $\partial W/\partial r$
term.

To derive the instability criterion, we consider the WKB solution 
of the form (valid for $kr\gg 1$)
\be 
W\propto \exp(i\sigma t+ikr).
\ee
Substituting into (\ref{warpw}), we obtain the dispersion relation
\be
\sigma=\left({3\nu_2\over 4r}k-\Omega_p\right)+i\left({1\over 2}
\nu_2 k^2-\Gamma_w\right).
\ee
For the instability to grow, we require ${\rm Im}(\sigma)<0$, or
\be
\Gamma_w>{1\over 2}\nu_2 k^2 \Longleftrightarrow
{\rm Instability}.
\ee
Since the radial wavelength is restricted to $\lambda\le r$, or
$k\ge 2\pi/r$, the instability criterion becomes
\be
\Gamma_w >{2\pi^2}{\nu_2\over r^2}\Longleftrightarrow
{\rm Instability}.
\label{criterion}\ee
This criterion has a simple physical interpretation: 
The magnetic torque drives the growth of disk tilt on 
a timescale $\Gamma_w^{-1}$, while viscosity tries to reduce
the tilt on a timescale $r^2/\nu_2$ (which is of the same order
as the radial drift time of the flow, $r/|V_r|$, if $\nu_2$ is
of the same order as $\nu_1$). Instability requires
$\Gamma_w^{-1}\lo r^2/\nu_2$.

Using (\ref{vr}) and (\ref{gammaw}), and assuming that $\nu_2/\nu_1$ is
independent of $r$, we can further reduce (\ref{criterion}) to
\be
r<r_w=\left({3\,\zeta\cos^2\theta\over 8\pi^2\cJ}{\nu_1\over\nu_2}
\right)^{2/7}\left({\mu^4\over GM\dot M^2}\right)^{1/7}.
\ee
This indicates that inside the critical {\it warping radius} $r_w$, 
the magnetic torque
can overcome the viscous force and make the disk tilt grow. Moreover,
with the expected\footnote{For Keplerian and inviscid (or very nearly so) 
disks, resonance between the epicyclic frequency and orbital frequency
leads to $\nu_2/\nu_1\simeq 1/(2\alpha^2)\gg 1$. However, the resonance is
delicate and might be destroyed by the effects of general relativity, 
self-gravity, magnetic fields and turbulence. See Ogilvie (1999) for
a discussion.}
$\zeta\sim 1$, $\nu_1/\nu_2\sim 1$ and $\cJ\le 1$,
the warping radius is of the same order of magnitude as the magnetosphere
radius $r_m$ (see eq.~[\ref{alfven}] and Appendix B for a discussion 
of the magnetosphere radius for arbitrary geometry).
Thus, as the disk approaches the magnetosphere, it will tend to 
be tilted with respect to the stellar spin even if at large radii
the disk normal is aligned with the spin axis.

Similar analysis can be applied to the case where
the disk is nearly antiparallel to the spin ($\pi-\beta\ll 1$). 
If we define $W\equiv (\pi-\beta){\rm e}^{i\gamma}$, then eq.~(\ref{warpw})
remains valid except the sign in front of $i\Omega_pW$ changes.
We obtain the same instability criterion as above. Thus, an antiparallel
disk tends to be driven toward a perpendicular configuration near the
magnetosphere radius. 

%=================================================================
\subsection{Magnetic Bardeen-Petterson Effect}

A tilted disk will be driven into precession by the torque
$\bN_{\rm prec}$ with ${\bf\Omega}_{\rm prec}=-\Omega_p\cos\beta\,\hato$
(see eq.~[\ref{omegap}]). What is the effect of this precession
on the disk tilt? An analogy can be made with the 
behavior of a disk undergoing Lense-Thirring precession around 
a rotating black hole (or any compact object)
(Bardeen \& Petterson 1975; Kumar \& Pringle 1985; Pringle 1992; 
Scheuer \& Feiler 1996). 

The gravitomagnetic force from the rotating object
(with spin angular momentum $\bJ=J\hato$)
drives the precession of the misaligned disk at angular frequency
${\bf\Omega}_{\rm LT}={2G\bJ/(c^2r^3)}$.
Bardeen \& Petterson (1975) pointed out that the action of
viscosity on the differentially precessing disk tends to align 
the rotation of the inner disk with the spin axis of the central object.
The Bardeen-Petterson radius $r_{\rm BP}$, that is, loosely speaking,
the radius inside which the disk aligns with the spin, is obtain by 
equating the precessional time $\Omega_{\rm LT}^{-1}$ and the
viscous time $r^2/\nu_2$ (Scheuer \& Feiler 1996), i.e.,
\be
r_{\rm BP}={2GJ\over c^2\nu_2}.
\ee
However, the transition from the warped outer disk to the aligned inner
disk is rather broad, and for this reason the timescale to 
achieve the Bardeen-Petterson alignment is much larger than the 
precession timescale at $r_{\rm BP}$ (Kumar \& Pringle 1985; Pringle 1992).
Numerical simulations (Pringle 1992) indicate that a steady state is
achieved on a time scale of order 
$(10-100)\,\Omega_{\rm LT}^{-1}$ (evaluated at $r_{\rm BP}$).

Now consider the effect of magnetically driven precession\footnote{In some
systems (such as accreting neutron stars with weak magnetic fields), 
the Lense-Thirring precession dominates (See \S 6). But here we neglect
$\Omega_{\rm LT}$.}. Setting $\Omega_p$ equal to $\nu_2/r^2$, we obtain 
the magnetic Bardeen-Petterson radius:
\be
r_{\rm MBP}=\left({3\sin^2\theta\over\pi\cJ D}{\nu_1\over\nu_2}\right)^{2/7}
\left({\mu^4\over GM\dot M^2}\right)^{1/7}.
\ee
Inside $r_{\rm MBP}$, the combined effect of viscosity and precession
tends to align the disk normal with the spin axis. 
We see that typically $r_{\rm MBP}$ is of the same order as $r_w$ (the
warping radius) and $r_m$
(the magnetosphere radius). Thus the precessional torque 
has an opposite effect on the
disk tilt as the warping torque discussed in \S 5.1. However, 
because of the broad warp-alignment transition expected for the magnetic
Bardeen-Petterson effect and the long timescale involved, we expect that the 
precession-induced alignment will be overwhelmed by the warping instability.

%\subsection{Comparison with the Radiation Driven Warping Instability}

%%%%%%%%%%%%%%%%%%%%%%%%%%%%%%%%%%%%%%%%%%%%%%%%%%%%%%%%%%%%%%%%%%%%%%
\section{EFFECT ON THE SPIN EVOLUTION}

How does the magnetically warped and precessing disk
affect the spin evolution of the central star? 
The angular momentum of the accreting gas is deposited at the magnetospheric
boundary and transferred to the the central object. In addition, 
there are back-reactions on the star associated with the magnetic torques on
the disk. Thus the spin angular velocity {\boldmath $\omega$} evolves
according to
\be
{d\over dt}\left(I\bo\right)=\dot M (GMr_m)^{1/2}\hatl
-\bcN_{\rm prec}-\bcN_{\rm warp},
\label{spin}\ee
where $I$ is the moment of inertia, 
$\bcN_{\rm prec}$ and $\bcN_{\rm warp}$
are the total warping and precessional torques acting on the disk
\footnote{In (\ref{spin}) we have neglected possible magnetic torque
[as in the Ghosh \& Lamb (1979a,b) picture] associated closed magnetic field
lines that thread the disk and connect the star. Such magnetic torque 
can be included by modifying the first term on the right-hand-side of
(\ref{spin}) to $\dot M\sqrt{GMr_m}\,f\,\hatl$, where $f$ is a dimensionless
function which depends on the ratio $\omega/\Omega(r_m)$ [See, e.g.,
eq.~(35) of Lai (1998) and references therein].}.
To calculate $\bcN_{\rm prec}$ and $\bcN_{\rm warp}$, we need to know 
the tilt angle $\beta$ as a function of $r$. In principle this can be
obtained by solving the tilt equation (\ref{warp}) or (\ref{warp1}).
But note that since the torques per unit area
(eqs.~[\ref{torque3b}]-[\ref{torque3c}]) are steep functions of $r$,
we can assume, as a first approximation, 
that $\beta$ is independent of $r$ near $r_m$. 
We then find
\be
\bcN_{\rm prec}=\int_{r_m}^\infty\!2\pi r \bN_{\rm prec}\, dr 
=-{4\mu^2\over 3\pi r_m^3}\sin^2\!\theta
\cos\beta\,(\hato\times\hatl),
\ee
and
\be
\bcN_{\rm warp}=\int_{r_m}^\infty\!2\pi r \bN_{\rm warp}\, dr 
=-{\zeta\,\mu^2\over 12 r_m^3}\cos^2\!\theta
\sin 2\beta\,\haty=-{\zeta\mu^2\over 6r_m^3}\cos^2\!\theta\cos\beta\,\left[
\hato-(\hato\cdot\hatl)\hatl\right].
\ee
Note that the ratio of the characteristic magnetic torque,
$\cN_{\rm mag}=\mu^2/r_m^3$, and the characteristic accretion torque,
$\cN_{\rm acc}=\dot M\sqrt{GMr_m}$, is 
\be
{\cN_{\rm mag}\over\cN_{\rm acc}}=\left({\mu^2\over\dot M\sqrt{GM}}\right)
{1\over r_m^{7/2}}=\eta^{-7/2},
\ee
[see eq.~(\ref{alfven}) and Appendix B].
The spin evolution equation can then be written as
\ba
{d\over dt}\left(I\bo\right)&=&\cN_{\rm acc}
\left[1+{\zeta\cos^2\!\theta\over 6\eta^{7/2}}\sin^2\!\beta\right]\cos\beta
\,\,\hato\nonumber\\
&&+\cN_{\rm acc}\left[{\zeta\cos^2\!\theta\over
6\eta^{7/2}}\cos^2\!\beta -1\right]\sin\beta\,\,\hato_\perp\nonumber\\
&&-\cN_{\rm acc}\left({4\over 3\pi\eta^{7/2}}\sin^2\!\theta\cos\beta
\right)\,\hatl\times\hato,
\label{spin2}\ea
where $\hato_\perp$ is a unit vector perpendicular to $\hato$ and lies
in the plane spaned by $\hatl$ and $\hato$ (see Fig.~1).
The physical effects of the three terms on the right-hand-side of
(\ref{spin2}) are evident: The first term is responsible to the spin-up
or spin-down (depending on the sign of
$\cos\beta$) of the star; the second term tends to
align or misalign the spin axis with the disk axis\footnote{Since the 
timescale to change $\hato$ is much longer than the timescale to
change $\hatl$, the change of the tilt angle $\beta$ is determined by 
the dynamics of $\hatl$ rather than $\hato$ (see eq.~[\ref{warp3}] and 
\S 4).},  and the third term 
induces precession of the star's spin axis $\hato$ around $\hatl$.
The timescales associated with these changes of $\bo$ (spin-up/spin-down,
alignment/misalignment and precession) are all of order
\be
\tau_{\rm spin}={I\omega\over\cN_{\rm acc}}
={I\omega \over \dot M\sqrt{GMr_m}}.
\label{tspin}\ee
This spin-changing timescale is typically much longer than the timescale
associated with changing the disk orientation (see \S 4 and \S6).

Note that eq.~(\ref{tspin}) represents the instantaneous spin-up/spin-down
timescale (i.e., at a given $\beta$). If the disk inclination 
$\beta$ wanders around $90^\circ$, we expect the secular spin-up of
the star to be slower (see \S 6).

%%%%%%%%%%%%%%%%%%%%%%%%%%%%%%%%%%%%%%%%%%%%%%%%%%%%%%%%%%%%%%%%%%%%%%
\section{ASTROPHYSICAL APPLICATIONS}

In this section we discuss/speculate several possible applications of
our theory. While preliminary, they demonstrate the potential importance
of the physical effects uncovered in previous sections. 

%===============================================================
\subsection{Spin Evolution of Disk-Fed X-Ray Pulsars}

Recent long-term, continuous monitoring of accreting X-ray pulsars with 
the BATSE instrument on the Compton Gamma Ray Observatory has revealed
a number of puzzling behaviors of the spins of these accreting, magnetized
neutron stars (see Bildsten et al.~1997 and references therein). 
Several well-measured disk-fed systems (e.g., Cen X-3, GX 1+4 and 4U 1626-67) 
display sudden transitions between episodes of steady
spin-up and spin-down, with the absolute values of spin torques approximately
equal (to within a factor of two).
Of special interest is the observed anticorrelation 
between the torque and X-ray luminosity during the spin-down phase of
GX 1+4 (i.e., the torque becomes more negative as the luminosity increases;
Chakrabarty et al.~1997). These features are at odds with previous
theoretical models, according to which the neutron star must be 
near spin-equilibrium in order to experience both spin-up and spin-down
(e.g., Ghosh \& Lamb 1979a,b;
see also Yi, Wheeler \& Vishniac 1997; Torkelsson 1998;
Lovelace et al.~1998). 
It has been noted that the observational data can be nicely explained if the
disk can somehow change its sense of rotation (Nelson et al.~1997) --- This
poses a significant theoretical problem, since the disk formed in a Roche lobe
overflow has a well-defined direction of rotation. It has been suggested
%that radiation-driven warping instability (identified by Pringle 1996)
%may give rise to the reversal of the inner region of the disk,
(Van Kerkwijk et al.~1998) that the disk reversal may be caused by 
the radiation-driven warping instability (Pringle 1996), although the extent
and timescale of the reversal remain unclear. (Recall that one expects
the radiation-driven instability to operate only at large radii; 
Pringle 1996.) Also, since the radiation driven warping does not 
directly depends on the spin orientation, it is not clear why the disk prefers
to wander around the perpendicular state ($\beta =90^\circ$).

Here we suggest that the magnetically driven warping instability 
uncovered in this paper plays an important role in the determining the spin 
behaviors of accreting X-ray pulsars. The magnetosphere is located at
\be
r_m=\eta\left({\mu^4\over GM\dot M^2}\right)^{1/7}
=\left(3.4\times 10^8\,{\rm cm}\right)\,\eta\,\,\mu_{30}^{4/7}M_{1.4}^{-1/7}
\dot M_{17}^{-2/7},
\label{alfven2}\ee
and we shall consider $\eta$ to be a constant of order unity
(see Appendix B).
Here $\mu_{30}$, $M_{1.4}$ and $\dot M_{17}$ are the neutron star's
magnetic moment, mass and accretion rate in units of $10^{30}$~G$\,$cm$^3$,
$1.4M_\odot$ and $10^{17}$~g~s$^{-1}$, respectively.
Near the magnetosphere, the disk lies in the so-called ``middle''
(gas pressure and scattering dominated)
region of the $\alpha$-disk solution (Shakura \& Sunyaev 1973; 
Novikov \& Thorne 1973). The surface density of the disk is
\be
\Sigma=\left(1.7\times 10^3\,{\rm g~cm}^{-2}\right)\,
\alpha^{-4/5}M_{1.4}^{1/5}\dot M_{17}^{3/5}
r_8^{-3/5}\cJ^{3/5},
\ee
where $r_8=r/(10^8~{\rm cm})$, and the possible effect of a threaded
magnetic field can be included in the definition of $\cJ$ (see Appendix A).
The growth rate of the magnetic warping instability (eq.~[\ref{gammaw}])
is given by
\ba
\Gamma_w &=&\left(0.035\,\,{\rm s}^{-1}\right)
\left(\zeta\cos^2\theta\right)\,\alpha^{4/5}
\mu_{30}^2\,M_{1.4}^{-7/10}\dot M_{17}^{-3/5}r_8^{-49/10}\cJ^{-3/5}\nonumber\\
&=&\left({1\over 5.3~{\rm day}}\right)
\left(\zeta\cos^2\theta\right)\,\left({\alpha\over 0.01}\right)^{4/5}
\mu_{30}^{-4/5}\dot M_{17}^{4/5}\left({r_m\over\eta r}\right)^{49/10}
\cJ^{-3/5}.
\label{gamma}\ea
As discussed in \S 2 and \S 4, the magnetic torque tends to drive 
the inner region (near the magnetosphere) of the disk toward a perpendicular
configuration ($\beta=90^\circ$) on a timescale of order $\Gamma_w^{-1}$.
Indeed, the $\beta=\pi/2$ state represents an ``attractor''.
In the idealized situation (each disk ring evolves independent of each
other), the tilt of a nearly orthogonal disk evolves according to
\be
{d\Delta\beta\over dt} =\Gamma_w\sin\beta\cos\beta\simeq -\Gamma_w\Delta\beta,
\ee
where $\Delta\beta\equiv\beta-\pi/2$ (see eq.~[\ref{warp3}]).
The star can then spin-up or spin-down, depending on the whether
$\beta<90^\circ$ or $\beta>90^\circ$ [see eq.~(\ref{spin2})]:
\be
I\dot\omega=\cN_{\rm acc}
\left[1+{\zeta\cos^2\!\theta\over 6\eta^{7/2}}\sin^2\!\beta\right]\cos\beta.
\ee
The characteristic spin-up/spin-down time scale (eq.~[\ref{tspin}]) is
\be
\tau_{\rm spin}=\left(7.9\times 10^3\,{\rm yrs}\right)\eta^{-1/2}
M_{1.4}^{-3/7}I_{45}\,\mu_{30}^{-2/7}\dot M_{17}^{-6/7}
\left({1\,{\rm s}\over P_s}\right),
\label{spinup}\ee
where $P_s$ is the spin period, and $I_{45}=I/(10^{45}\,{\rm g~cm}^2)$.

In our picture, the observed sign switching of $\dot\omega$ in several X-ray
pulsars is associated with the ``wandering'' of $\beta$ around the
``preferred'' value ($\beta=90^\circ$). Such ``wandering'' needs not be
periodic:  Consider a disk initially at $\beta=90^\circ$. Imagine that a 
perturbation in the accretion induces a negative (not necessarily small)
$\Delta\beta=\beta-90^\circ$ --- This is achieved on a viscous
timescale (which is of the same order of magnitude as $\Gamma_w^{-1}$
at the inner disk edge; see \S 4.1). The star spins up. The magnetic torque
then drives $\beta$ toward $90^\circ$ on the timescale of $\Gamma_w^{-1}$,
at which point another perturbation (which cannot be faster than
$\Gamma_w^{-1}$) can induce another $\Delta\beta$, which can be
either negative or positive, and the star will then continue to spin-up or 
switch to spin-down. We expect that the timescale of the switching 
between spin-up and spin-down is of order a few times 
$\Gamma_w^{-1}$ (evaluated at the
inner disk boundary). For $\alpha\sim 0.1-0.01$ and $\dot M_{17}\sim 0.1$,
$\mu_{30}\sim 1$ (typical of X-ray pulsars), eq.~(\ref{gamma}) gives
$\Gamma_w^{-1}\sim 5-30$~days (for $\zeta\cos^2\!\theta=1$),
comparable to what is observed in Cen X-3 ($P_s=4.8$~s) 
(Bildsten et al.~1997). For 4U 1626-67 ($P_s=7.6$) and GX 1+4 ($P_s=120$~s),
the sign of $\dot\omega$ switchs once in $10-20$~yr. Since equation 
(\ref{gamma}) is uncertain and depends on many parameters, it is
conceivable that such long switching time can be accommodated
(e.g., with $\zeta\cos^2\!\theta=0.1$, $\alpha=0.01$,
$\dot M_{17}=0.1$ and $\mu_{30}=10$, we find $\Gamma_w^{-1}\simeq
6$~years).

Note that eq.~(\ref{spinup}) represents the ``instantaneous'' 
spin-up/spin-down time of the neutron star. Since the outer disk has 
a well-defined direction (presumably with $\beta<90^\circ$), 
the inner disk will unlikely spend equal amount of time 
in the prograde ($\beta<90^\circ$) phase and the retrograde 
($\beta>90^\circ$) phase. Thus we expect that on a longer time scale
(longer than the disk reversal time $\Gamma_w^{-1}$), the neutron
star will experience secular spin-up. This expectation is borne out by
observations, although it is difficult to predict the long-term spin-up
rate.   

A full study or simulation of the nonlinear behavior of
the inner disk tilt would be desirable to make more meaningful
comparison with observations. But our discussion and estimate
given above indicates that magnetically driven warping may be a crucial
ingredient in explaining the spin behaviors of disk-fed X-ray pulsars.
Other physical effects (such as radiation driven warping and propeller 
effect; Van Kerkwijk et al.~1998, Lovelace et al.~1998) may also play a role.

%================================================================
\subsection{Weakly Magnetized Neutron Stars: Quasi-Periodic Oscillations}

Rapid variability in low-mass X-ray binaries (LMXBs), containing
weakly magnetized ($B\sim 10^7-10^{9}$~G) neutron stars, 
has been studied since the discovery of the so-called 
horizontal-branch oscillations (HBOs)
in a subclass of LMXBs called Z sources
(van der Klis et al.~1985; Hasinger \& van der Klis 1989). 
The HBOs are quasi-periodic oscillations (QPOs) (with the $Q$-value 
$\nu/\Delta\nu$ of order a few, and rms amplitude $\lo 10\%$)
which manifest as broad Lorentzian peaks in the X-ray power spectra 
with centroid frequencies in the range of $15-60$~Hz which are positively
correlated with the inferred mass accretion rate (see van der Klis 1995 for a
review). For many years, the standard interpretation for the HBOs 
has been based on the magnetosphere beat-frequency model, first 
advocated by Alpar \& Shaham (1985) and Lamb et al.~(1985), in which the HBO
is identified  with the difference frequency between the Keplerian frequency
at the magnetospheric boundary and the spin frequency of the
neutron star (see Ghosh \& Lamb 1992 for a review). However,
recent observations of kHz QPOs ($500-1200$~Hz) in at least eighteen
LMXBs by the Rossi X-ray Timing Explorer (RXTE) have called into question
this interpretation of HBOs (see van der Klis 1998a,b
for review). The kHz QPOs (with $Q$ up to $100$ and rms amplitude up to 
$20\%$) often come in pairs, and both frequencies move up and down
as a function of photon count rate, with the separation frequency roughly
constant (The clear exceptions are Sco X-1 and 4U 1608-52;
van der Klis et al.~1997, Mendez et al.~1998a). In most Z sources,
the 15-60 Hz HBOs appear simultaneously with 
the kHz QPOs, while in several atoll sources (which are thought to have weaker
magnetic fields and smaller accretion rates than the Z sources), 
broad peaks at %$20-40$~Hz 
$10-50$~Hz in the power spectra (similar to the HBOs in the Z sources) have
also been detected at the same time when the kHz QPOs appear. While the origin
of the kHz QPOs is uncertain, it is natural to associate the higher frequency
QPO with the orbital motion at the inner edge (perhaps the magnetosphere
boundary) of the accretion disk, and the lower-frequency QPO may result
from the (perhaps imperfect) beat between the Kepler frequency and the neutron
star spin frequency\footnote{See Stella \& Vietri (1999) for an alternative
interpretation of the lower QPO peak which does not involve beating.} 
--- This beat frequency interpretation is supported
by the observations of a third, nearly coherent QPO which occurs
during X-ray bursts, at a frequency approximately equal to the frequency
difference between the twin kHz peaks or twice that value. (An exception is 4U
1636-53, Mendez et al.~1998b.) Clearly, if this generic identification of
kHz QPOs is correct, the beat between the $\sim 1$~KHz Keplerian frequency
at the disk inner edge and the $\sim 300$~Hz spin frequency cannot produce
the $10-60$~Hz low-frequency QPOs (LFQPOs: HBOs in the Z 
sources and similar features in the atoll sources), 
unless one postulates (Miller, Lamb \&
Psaltis 1998) that the inner disk edge lies inside
the magnetosphere and is unaffected by the magnetic field.
%This is rather unlikely in the author's opinion.

Stella and Vietri (1998) suggested that the $10-60$~Hz LFQPOs
are associated with Lense-Thirring precession of the inner accretion 
disk around the rotating neutron star. Assuming that the low-frequency QPO and
the kHz QPOs are generated at the same radius in the disk, one obtains,
to leading order, $\Omega_{\rm LT}=(2I\omega/Mc^2)\Omega^2$ (where 
$\Omega,~\Omega_{\rm LT}$ and $\omega$ are the orbital, Lense-Thirring, and
spin angular frequencies, respectively). Thus the LFQPO frequency 
depends quadratically on the kHz QPO frequency. (The classical precession
associated with spin-induced oblateness of the star, as well as
relativistic effect beyond the Lense-Thirring formula, both can introduce
small correction to this correlation; see
Stella \& Vietri 1998, Morsink \& Stella 1999.)
This is consistent with observations of several sources
(e.g., the Z sources GX17+2, GX5-1, Sco X-1, and the atoll source
4U1728-34; see Ford \& van der Klis 1998,
Psaltis et al.~1999 and references therein)
\footnote{Observations indicate that the ratio $I/M$ required to
fit the expected $\Omega_{\rm LT}-\Omega$ relation is a factor of $2-4$
larger than allowed by neutron star equation of state. This situation 
can be improved if the observed LFQPO frequency is the second harmonic of the
fundamental precession frequency (See, e.g., Stella \& Vietri 1998, 
Morsink \& Stella 1999, Psaltis et al.~1999). 
In some Z sources, one requires that the observed HBO frequency
is four times the fundamental precession frequency in order to produce
reasonable $I/M$. Alternatively, the spin frequency is twice of what is
inferred from the difference between the twin kHz QPOs ---
This would make the beat frequency interpretation of the lower kHz peak 
invalid. It would be interesting to search for
``sub-harmonic'' feature of the HBOs in the power spectra 
(see Ford \& van der Klis 1998 for possible evidence of such a sub-harmonic
feature in 4U 1728-34).}.
%Even if one adopt the (reasonable) assumption that the HBOs correspond to 
%the second harmonic of the fundamental precession frequency, observations
%of the Z sources still indicate that the required $I/M$ is larger than allowed
%by nuclear equation of state (unless the spin frequency is twice of what is
%inferred from the difference between the twin kHz QPOs); see Psaltis et
%al.~1998. Perhaps a more detailed calculation of $\nu_{\rm prec}$ (see 
%discussion at the end of this subsection) can
%help to resolve this puzzle.} 

For the Lense-Thirring interpretation of HBOs to be viable, the inner disk 
must be tilted with respect to the stellar spin axis. The Bardeen-Petterson
effect tends to keep the inner region of the disk [inside $r_{\rm BP}$,
typically at $(100-1000)GM/c^2$] co-planar with the star 
(Bardeen \& Petterson 1975). Radiation driven warping (Pringle 1996)
is only effective at large disk radii. While global disk warping modes
may exist with nonzero tilt near the inner disk boundary (Ipser 1996;
Markovi\'c \& Lamb 1998),
%\footnote{It is not clear that in inner disk 
%boundary conditions have been treated properly in these calculations.},
an external driving force is needed to excite them. 

Here we suggest that the magnetic warping torque provides 
a natural driver for the disk tilt near the inner accretion disk
\footnote{Vietri and Stella (1998) suggested that if the accretion disk is
inhomogeneous, diamagnetic blobs can be lifted above the equatorial plane
through resonant interaction with the star's magnetic field near the 
the corotation radius. See Appendix C.}. For typical parameters
of LMXBs, the magnetosphere is located at\footnote{Note that since $r_m$ is
so close to the inner-most stable orbit, general relativistic effect 
tends move the inner disk edge to a radius larger than what is given in
(\ref{alf2}); see Lai (1998). Here we shall neglect such complication.}
\be
r_m=\left(18\,{\rm km}\right)\,\eta\,\mu_{26}^{4/7}M_{1.4}^{-1/7}\dot
M_{17}^{-2/7}
\label{alf2}\ee
where $\mu_{26}=\mu/(10^{26}\,{\rm G~cm}^3)$. 
Assuming that near the magnetosphere the disk is described by the ``inner
region'' (radiation and scattering dominated) solution 
of $\alpha$-disk (Shakura \& Sunyaev 1973; Novikov \&
Thorne 1973), we have for the disk surface density and half-thickness
\ba
\Sigma &=&\left(105\,\,{\rm g~cm}^{-2}\right)
\alpha^{-1}M_{1.4}^{-1/2}\dot M_{17}^{-1}r_6^{3/2}\cJ^{-1},\label{surf}\\
H&=& \left(1.1\,{\rm km}\right)\dot M_{17}\,\cJ,\label{height}
\ea
where $r_6=r/(10^6\,{\rm cm})$.
The growth rate of the magnetic warping instability (eq.~[\ref{gammaw}])
is then
\ba
\Gamma_w &=&\left(555\,{\rm s}^{-1}\right)\left(\zeta\cos^2\!\theta\right)
\alpha\,\mu_{26}^2\dot M_{17}\,r_6^{-7}\cJ\nonumber\\
&=&\left(1.0\,\,{\rm s}^{-1}\right)\left(\zeta\cos^2\!\theta\right) 
\left({\alpha\over 0.1}\right)M_{1.4}\,\mu_{26}^{-2}\dot M_{17}^3
\left({r_m\over\eta r}\right)^7\cJ.
\ea
The magnetic effect also results in a precessional torque on the disk. 
From eq.~(\ref{prec3}) and using (\ref{surf})-(\ref{height}), we find that 
the precession frequency associated with this magnetic torque is
\ba
\nu_{\rm prec}&=&-\left(0.21\,{\rm Hz}\right)
\left(\sin^2\!\theta\cos\beta\right)\left({\alpha\over 0.1}\right)
M_{1.4}\,\mu_{26}^{-2}\dot M_{17}^3\left({r_m\over\eta r}
\right)^7 D^{-1}\cJ\nonumber\\
&=&-\left(0.60\,{\rm Hz}\right)\left(\sin^2\!\theta\cos\beta\right)
\left({\alpha\over 0.1}\right)M_{1.4}^{3/7}\mu_{26}^{-12/7}
\dot M_{17}^{20/7}\eta^{1/2}\left({r_m\over\eta r}\right)^7 \cJ^{1/2},
\label{precc}\ea
where in the second equality we have used $D=(2H/r_m)^{1/2}$
[see eq.~(\ref{Dfunc})].
%While our ignorance about the magnetosphere boundary layer
%precludes a more accurate calculation, 
This should be compared with the Lense-Thirring precession frequency
\ba
\nu_{\rm LT}&=&\left(44.4\,{\rm Hz}\right)\,I_{45}\,r_6^{-3}\left({\nu_s
\over 300\,{\rm Hz}}\right)\nonumber\\
&=&\left(8.1\,{\rm Hz}\right)M_{1.4}^{3/7}\,I_{45}\,\mu_{26}^{-12/7}\dot
M_{17}^{6/7}\left({\nu_s\over 300\,{\rm Hz}}\right)\left({r_m\over
\eta r}\right)^3,
\ea
where $\nu_s$ is the spin frequency.
Note that for $\beta<90^\circ$ 
the magnetically driven precession is opposite to 
the stellar spin axis (thus the negative sign in eq.~[\ref{precc}]),
while Lense-Thirring precession is in the same direction as the spin.
For a given source, the magnitude of $|\nu_{\rm prec}|$ is typically smaller
than $\nu_{\rm LT}$, but it becomes increasingly important with 
increasing $\dot M$ (since $\nu_{\rm LT}\propto\dot M^{6/7}$ while
$\nu_{\rm prec}\propto\dot M^{20/7}$).
This may explain 
the observed flattening of the correlation between the LFQPO frequency 
and kHz QPO frequency as the latter increases
\footnote{Since the classical precession rate due to the oblateness of the 
star is negative (for $\cos\beta>0$) and scales as $r^{-7/2}$
(while $\nu_{\rm LT}\propto r^{-3}$)
(Stella \& Vietri 1998, Morsink \& Stella 1999),
it may also explain this observed flattening, provided that 
the spin frequency is much higher than inferred from the kHz QPOs
and the burst QPOs.
However, because of the similar power-law indices ($r^{-7/2}$ vs. $r^{-3}$), 
it is difficult to explain why the $\nu_{\rm LFQPO}\propto
\nu_{\rm kHzQPO}^2$ scaling breaks down only at very high 
kHz QPO frequency (Psaltis et al.~1999).}.

We note that although our qualitative conclusion that magnetic effect
can induce tilt of the inner accretion disk and therefore facilitate
its precession is robust, the analytical expressions given in this 
subsection only serve as an order-of-magnitude estimate which
indicates the potential importance of the magnetically driven precession.
More detailed calculations (including global mode analysis) 
are needed to make more meaningful comparison
with observational data. In addition to the intrinsic uncertainties
associated with the $\alpha$-disk model and the magnetosphere boundary layer,
there are two complications that may prove important for LMXBs: (i) 
General relativistic effect can modify the inner radius of the disk so that
it is not just determined by magnetic-plasma stress balance (Lai 1998);
(ii) The magnetic field is not expected to be dipolar. This comes
about either because of the multipole fields from the central star, or,
even if the intrinsic stellar field is dipolar, the (partially diamagnetic)
disk can enhance the field in the boundary layer (Lai, Lovelace \& Wasserman 
1999). 

%At this point, the association between disk precession and 
%low-frequency QPOs must be considered tentative (see footnote 8). 
%Even if the association turns out to be non-existent, 
%it would be interesting to understand why the precessional 
%frequency does not show up in the X-ray power spectra 
%(given that disk tilting is inevitable from our theoretical consideration). 

%============================================================
\subsection{Super-Orbital Periods in X-ray Binaries}

Here and in \S 6.4 we speculate upon two additional applications 
of our theory.

Many X-ray binaries are known to exhibit long-term cycles in 
their X-ray or optical luminosities (see Priedhorsky \& Holt 1987 and
references therein).  Of particular interest is the
well-established ``third'' period (longer than
the orbital period) in Her X-1 (35 d) (Tananbaum et al.~1972),
LMC X-4 (30.5 d) (Lang et al.~1981) and
SS433 (164 d) (Margon 1984). It is generally thought that
these super-orbital periods result from the precession of a
tilted accretion disk. For Her X-1, Katz (1973) proposed that the precession
was forced by the torque from the companion star, but left unexplained the
origin of the disk's misalignment with respect to the orbital plane. 
Recently it has been suggested that the radiation-driven warping
instability (Pringle 1996) is responsible for producing warped, precessing
disks in many X-ray binaries which exhibit long-term cycles (Maloney
\& Begelman 1997; Wijers \& Pringle 1998). 

It is likely that magnetic torque plays a role in driving 
the warping of the inner disk. The observed 
systematic variation of the X-ray pulse profile of Her X-1 requires
the inner edge of the disk to be significantly warped 
(Sheffer et al.~1992; Deeter et al.~1998 and references therein). In addition, 
the magnetically driven precession frequency is
\be
\nu_{\rm prec}=-\left({1\over 26\,{\rm day}}\right)
\left(\cos\beta\sin^2\!\theta\right)\,\left({\alpha\over 0.01}\right)^{4/5}
\mu_{30}^{-4/5}\dot M_{17}^{4/5}\left({r_m\over\eta r}\right)^{49/10}
\cJ^{-3/5}D^{-1},
\label{precf}\ee
where $r_m$ is the magnetosphere radius [eq.~(\ref{alfven2})], and
we have adopted the ``middle region'' solution of the $\alpha$-disk
(near the inner edge of the disk, $D\sim 0.2$). 
This is comparable to the observed super-orbital periods. Moreover,
the magnetically driven precession is retrograde with respect to the 
direction of rotation of the disk, in agreement with observations. 
Of course, the precession rate is a function of $r$ and $\beta$,
so a modal analysis is needed to determine the global precession
period. 

%============================================================
\subsection{T Tauri Stars}

It is well established that classical T Tauri stars (CTTS) have circumstellar
disks; Evidence also exists for magnetospheric accretion induced by the
stellar magnetic field (e.g., Hartmann 1998).  
For typical parameters ($M\simeq 0.5\,M_\odot$, $R\simeq 2R_\odot$, $\dot
M=10^{-9}-10^{-7}\,M_\odot$~yr$^{-1}$, and surface field $B_\star\sim 1$~kG),
the magnetosphere is located at a few stellar radii. 
Using $\mu=B_\star R^3$, we find that the growth time for the warping
instability at the inner region of the disk 
is given by [see eq.~(\ref{gammaw})]
\be
\Gamma_w^{-1}=(6.5\,{\rm days})\left({1\,{\rm kG}\over B_\star}\right)^2\!
\left({2R_\odot\over R}\right)^6\!\left({M\over 0.5M_\odot}\right)^{1/2}\!\!
\left({r\over 8R_\odot}\right)^{11/2}\!\!
\left({\Sigma\over 1\,{\rm g\,cm}^{-2}}\right)\!
\left(\zeta\cos^2\theta\right)^{-1}.
\ee
The precession period of the tilted disk (see eq.~[\ref{prec3}])
is of the same order of magnitude as $\Gamma_w^{-1}$. The surface density
$\Sigma$ is unknown, but reasonable estimates give $\Sigma\sim
1-100$~g~cm$^{-2}$ (Hartmann 1998). Therefore $\Gamma_w^{-1}$ ranges from 
days to years, much shorter than the lifetime of
T Tauris stars ($\sim 10^6$~years). 

It has been observed that the photometric periods ($5-10$~days)
of CTTS vary (by as much as $30\%$) on a timescale of weeks (Bouvier et
al.~1995). The origin of this variability is unknown. If we interprate
the photometric period as the orbital period at the magnetosphere
boundary (Bouvier et al.~1995), then we may understand the period 
variation in the context of warped, precessing disks: As the inner disk 
warps and precesses, it experiences different stellar magnetic field, 
and thus the inner disk radius varies. 

It is also of interest to consider the effect of magnetically driven 
warping on the rotation of T Tauri stars.
The projected rotation velocity of CTTS with masses 
$M\lo 1\,M_\odot$ is about $20$~km~s$^{-1}$, only $10\%$ of
the breakup speed (e.g., Bertout 1989). This is at odds with the expectation
that T Tauri stars are formed by the gravitational collapse
of rotating molecular cores and the presence of disks 
surrounding the stars. Theories which explain the slow rotations generally
invoke the interaction between the disk and the stellar magnetic field of 
a few kG (e.g., K\"onigl 1991; Cameron \& Campbell 1993; Shu et al.~1994; 
Yi 1995; Armitage \& Clarke 1996). 
Since the growth time for disk warping is short,
we may expect the inner disk of T Tauri stars to wander around the
``preferred'' perpendicular state. The star therefore experiences both spin-up
and spin-down during its evolution, analogous to the behavior of X-ray pulsars
(see \S 6.1). The net, secular spin-up rate is expected to be much smaller 
than that based on the canonical spin-up torque ($\sim \dot M\sqrt{GMr_m}$).

\section{CONCLUSION}

In this paper we have identified a new magnetically driven warping 
instability which occurs in the inner accretion disk
of a magnetized star (\S 2 and \S 4). 
Despite the uncertainties in our understanding
of the magnetosphere--disk interactions (particularly the global magnetic 
field structure), the existence of
the instability seems robust, and requires that some vertical field lines 
(either from the star or intrinsic to the disk) thread the disk
and get twisted by the disk rotation. The general consequence of
the instability is that the normal vector of the inner disk (near 
the magnetosphere) will be tilted with respect to the stellar spin axis.
We have also shown that the disk can be driven magnetically into
precession around the spin axis due to the interaction between
the screening surface current and the stellar magnetic field (\S 2). 
In addition, certain regions of the disk are subjected to 
resonant magnetic forces which may affect the structure and dynamics
of the disk (\S 3). These magnetic effects on accretion disks have largely 
been overlooked in previous studies of accretion onto magnetic stars
\footnote{We note that the effects studied in this paper are
quite different from the situation considered by Agapitou et al.~(1997),
who studied the bending instability in the inner (sub-Keplerian) disk 
which corotates with the star (with the magnetic axis aligned with the spin
axis). Such a disk may or may not exist (see Spruit \& Taam 1990).}. 

We have applied our theory to several different types of astrophysical
systems, including X-ray pulsars, low-mass X-ray binaries, and T Tauri stars
(\S 6). These applications should be considered preliminary,
but they indicate that the magnetically driven disk warping and precession
can potentially play an important role in determining the
observational behaviors of these systems. Of particular interest is that 
the magnetically warped inner disk may provide a natural explanation for the 
longstanding puzzle of torque reversal as observed in a number of X-ray 
pulsars. Also, the tilted disk may be responsible for the
rich phenomenology of time variability (such as QPOs) observed in 
weakly magnetized accreting neutron stars in LMXBs. 

Much work is needed to understand better the effects studied in this paper and 
their observational manifestations in different astrophysical systems. In our
analysis, we have intentionally avoided (or bypassed),
by using parametrized models, the uncertainties
associated with magnetosphere -- disk interactions (see, e.g., Appendix A,B),
but clearly the study of disk warping and precession
based on more specific models (with or without outflows)
would be useful. The role of intrinsic disk field needs to be examined
further (see \S 2.4). There remains uncertainty in the description of
nonlinear warped disks, and a full numerical simulation of the 
nonlinear development of the warping instability would be valuable. 
In the case of low-mass X-ray binaries, the effects of complex
magnetic field topology (other than dipole) and general relativity
should be included to access the QPO phenomenology (see \S 6.2). More detailed
comparison with observational data will be useful. 
The role of magnetically driven resonances need to be studied further to
determine whether they will produce any observable features. 
We hope to address some of these issues in the future.

\acknowledgments

I thank Richard Lovelace, Phil Maloney,
Dimitrios Psaltis, Marten van Kerkwijk and Ethan
Vishniac for useful discussion/comment. I also thank Brad
Hansen for informing me of some important references. 
I became puzzled by the spin-up/spin-down behavior of X-ray pulsars
several years ago at Caltech through discussions with Lars Bildsten,
Deepto Chakrabarty, Rob Nelson and Tom Prince. 
I acknowledge the support from a NASA ATP grant and 
a research fellowship from the Alfred P. Sloan foundation, 
as well as support from Cornell University.

\appendix

\section{MAGNETIZED ACCRETION DISK}

For an accretion disk threaded by a vertical magnetic field
$B_z$, the steady-state angular momentum equation reads:
\be
V_r{dl\over dr}={r\over\Sigma}\left[{1\over r^2}{d\over dr}\left(
r^3\Sigma\nu_1{d\Omega\over dr}\right)+{B_z\Delta B_\phi\over 2\pi}\right],
\ee
where $l$ is the angular momentum per unit mass, $\Sigma$ is the surface
density, $\Delta B_\phi$ is the $\phi$-component of the magnetic field
produced by twisting $B_z$ and evaluated at the upper disk surface.
Integrating over $r$ and using the mass continuity equation
$\dot M=-2\pi r\Sigma V_r$, we obtain the conservation equation for angular
momentum:
\be
\dot M l_0=\dot M l+2\pi\nu_1 r^3\Sigma\Omega'+\dot M l_B,
\label{angu}\ee
where $l_0$ is a constant, $\Omega'\equiv d\Omega/dr$, and
\be
\dot M l_B=-\int_r^\infty\!\!dr\,r^2B_z\Delta B_\phi.
\label{angu1}\ee
Equation (\ref{angu}) says that the rate of net angular momentum
transfer through the disk, $\dot M l_0$, is equal to the sum of
the rates of advective, viscous, and magnetic transport.
We can rewrite (\ref{angu}) as
\be
\Sigma=-{\dot M\Omega\over 2\pi r\nu_1\Omega'}\,\cJ,
\label{sig}\ee
where 
\be
\cJ\equiv 1-{l_0-l_B\over l}.
\ee
The radial velocity is
\be
V_r={\nu_1\Omega'\over\Omega}\cJ^{-1}.
\label{vr1}\ee
For a Keplerian flow, $\Omega=\sqrt{GM/r^3}$, $l=\sqrt{GMr}$, and
eqs.~(\ref{sig})-(\ref{vr1}) reduce to (\ref{vr}).
The standard thin disk equations are recovered if we set $l_B=0$.

The above equations are quite general, but the actual expression for $l_B$
depends on the behavior of the disk magnetic field, and thus should be
viewed as being uncertain. Consider a specific ansatz:
\be 
B_z=B_0\left({R\over r}\right)^3,~~~~\Delta B_\phi=-\zeta B_z
\ee
(see eqs.~[\ref{bphi3}]-[\ref{bz3}]), where 
$B_0$ measures the magnetic field at the stellar surface ($r=R$).
Assuming that $\zeta$ is a constant, we find
\be
l_B={\zeta B_0^2\over 3\dot M r^3}={1\over 3}\,b^2l_R\left({R\over r}\right)^3,
\ee
where $l_R\equiv\sqrt{GMR}$ and $b^2\equiv \zeta B_0^2R^3/(\dot M l_R)$ is
dimensionless. More general expressions (including spin dependence,
multipole field and general relativistic effect) can be found in Lai (1998).
The inner edge of the disk, $r_m$, can be formally defined as where where 
$d(l+l_B)/dr=0$; we find
\be 
r_m=R\,(2b^2)^{2/7}=\left({2\zeta B_0^2R^6\over\dot M\sqrt{GM}}\right)^{2/7}
\ee
(cf.~eq.~[\ref{alfven}]).
We then have
\be
l_0=l(r_m)+l_B(r_m)={7\over 6}\,l(r_m)={7\over 6}\sqrt{GMr_m},
\ee
and
\be
\cJ=1-{1\over 6}\left({r_m\over r}\right)^{1/2}\left[7-\left({r_m\over
r}\right)^3\right].
\ee
This would suggest $\cJ=0$ at $r=r_m$. But of course as $r$ approaches
the inner disk boundary, this expression breaks down, as the inertial 
term of the radial equation must be taken into account (see Lai 1998
for a magnetic slim disk model). 

%==========================================================
\section{MAGNETOSPHERE RADIUS FOR GENERAL FIELD GEOMETRY}

In the main text, we have intentionally avoided the precise
definition of the magnetosphere radius $r_m$ for general orientations
of $\hatl$,~$\hato$ and $\hatmu$. This is because the exact determination 
of $r_m$ requires a detailed model of the structure of magnetosphere--disk
boundary layer, for which no definitive theory exists
(useful discussions are contained in the references cited in \S 1). 
We offer the following three possibilities. They should be viewed
only as an educated guess, although they all give a scaling relation as in eq.~(\ref{alfven}). 

(i) {\it Ansatz 1}: If the static vertical field
$B_z=-(\mu/r^3)\cos\beta\cos\theta$ threads the disk (see \S 2.3),
it will affect the angular momentum transport in the disk through
the magnetic stress $\propto B_z\Delta B_\phi$, where $B_\phi=-\zeta B_z$
is the field created by winding $B_z$
(see eqs.~[\ref{angu}]-[\ref{angu1}]). When the condition
\be
\dot M {dl\over dr}+r^2B_z\Delta B_\phi=0
\label{mag1}\ee
(where $l=\sqrt{GMr}$ is the angular momentum per unit mass) is satisfied,
no viscosity is needed to induce accretion.  We may thus use this condition 
to determine the inner edge of the Keplerian disk, giving
\be
r_m=\left(2\zeta\cos^2\!\beta\cos^2\!\theta\right)^{2/7}
\left({\mu^4\over GM\dot M^2}\right)^{1/7}.
\ee

(ii) {\it Ansatz 2}: Here we shall follow the consideration similar to
that in Arons (1993). Assume that the disk is largely diamagnetic, 
but strong dissipation exists at the boundary layer, where plasma stress
and magnetic stress balance, i.e.,
\be
\rho V_rV_\phi={B_z\Delta B_\phi\over 4\pi}.
\label{balan}\ee
The vertical magnetic field at the disk inner edge ($r=r_m$) is given by
\be
B_z(r_m)=-{4\mu\over\pi r_m^3}\left({r_m\over 2H}\right)^{1/2}\cos\chi
\ee
(Aly 1980). 
Shear in the boundary layer induces $\Delta B_\phi=-\zeta_B B_z$ with 
$\zeta_B\lo 1$. Using $V_\phi=\sqrt{GM/r}$ and $V_r=-\dot M/(4\pi r\rho H)$,
we find
\be
r_m=\left({8\,\zeta_B\over\pi^2}\cos^2\!\chi\right)^{2/7}
\left({\mu^4\over GM\dot M^2}\right)^{1/7}.
\label{mag2}\ee
Since $\cos\chi=\cos\beta\cos\theta-\sin\beta\sin\theta\sin\omega t$,
eq.~(\ref{mag2}) implies that the magnetosphere boundary is modulated 
at the spin frequency $\omega$. If we average the magnetic stress over time,
we may replace $\cos^2\!\chi$ by $[\cos^2\!\beta\cos^2\!\theta
+(1/2)\sin^2\!\beta\sin^2\!\theta]$.

(iii) {\it Ansatz 3}: This is similar to (ii), except that we now assume that 
the static vertical field threads the disk [see eq.~(\ref{bz0})],
so that at $r=r_m$,
\be
B_z(r_m)=-{\mu\over r_m^3}\cos\beta\cos\theta+{4\mu\over \pi r_m^3}
\left({r_m\over 2H}\right)^{1/2}\sin\beta\sin\theta\sin\omega t.
\ee
The condition (\ref{balan}) then gives
\be
r_m=\left({8\,\zeta_B\over\pi^2}\right)^{2/7}
\left[\sin\beta\sin\theta\sin\omega t-
{\pi\over 4}\left({2H\over
r_m}\right)^{1/2}\!\!\cos\beta\cos\theta\right]^{4/7}
\left({\mu^4\over GM\dot M^2}\right)^{1/7}.
\label{mag3}\ee

We see that in general, the inclined rotating dipole will 
produce a time-varying magnetosphere boundary. This variability
may be in addition to that associated the winding-up and reconnection
of field lines (see the end of \S 2.4). 
%It is of interest to explore the observational consequences 

%==========================================================
\section{MOTION OF DIAMAGNETIC BLOBS}

It has recently been suggested that under certain conditions, 
the accretion disk onto magnetized object may
consist of diamagnetic blobs (King 1993; Vietri \& Stella 1998).
Here we study how the stellar magnetic field affects
the motion of diamagnetic blobs which lie in circular orbits
of the disk plane. Each blob moves independent of each other, 
and is subjected to the gravitational force from the central star
and the magnetic drag force which arises when it moves across the field 
lines (see below). It is not clear that a realistic disk will behave
as a collection of individual blobs, nor is it clear that 
the blob can survive for a long time (e.g., the blobs may be 
subjected further Kelvin-Helmhotz instability which tends to
break up the blobs) --- These issues are 
not addressed in this paper. Our purpose here is to understand the
dynamics of the blob under the afore-mentioned assumptions. 
 
Vietri and Stella (1998) has studied some aspects of this problem.
They assumed that the spin axis is aligned with the angular momentum axis of
the orbiting blob. They 
%considered the vertical magnetic force on the blob, and
showed that vertical resonances exist in the region where
$\omega/2\le\Omega\le 3\omega/2$ (where $\omega,\,\Omega$
are the spin frequency and orbital frequency),
and suggested that the resonances can pump the blobs out of the
equatorial plane. 
Here we consider the more general case where the 
orbital angular momentum of the blob is misaligned with 
the spin axis. We show that even without the resonances,
there exists a magnetic torque which can 
induce tilt of the orbit of the blob. 

Our setup is given in Fig.~1. The magnetic drag force on the blob
results from the Lorentz force on the screening current on the blob's
surface. The blob loses energy by exciting Alf\'ven waves in the surrounding
plasma. The drag force 
per unit mass is given by
\be
\bF_{\rm drag}=-{\Delta \bV_\perp\over\tau_d},
\ee
and the drag time scale is 
\be
\tau_d={c_Am\over B^2d^2}.
\ee
Here $c_A$ is the Alf\'ven speed in the interblob plasma (where the field
strength is $B$), and $m$ and $d$ is the mass and characteristic
size of the blob (Drell, Foley \& Ruderman 1965). 
We shall assume that the interblob field is that of the stellar dipole
(thus neglecting possible screening due to the blobs):
\ba
B_r&=&{2\mu\over r^3}\sin\chi\cos(\phi-\phi_\mu),\label{br4}\\
B_\phi&=&{\mu\over r^3}\sin\chi\sin(\phi-\phi_\mu),\label{bphi4}\\
B_z&=&-{\mu\over r^3}\cos\chi.\label{bz4}
\ea
The relative velocity between the blob and the field line 
[at location $(r,\phi,z=0$)] is
\be
\Delta\bV=\Omega r\,\hat\phi-\Omega_s\hato\times\br.
\ee
The projected relative velocity perpendicular to the field line is then
\ba
\Delta\bV_\perp &=&
\Delta\bV-{(\bB\cdot\Delta\bV)\,\bB\over |\bB|^2}\nonumber\\
&=& \Delta\Omega r\,\hat\phi
+\omega r\sin\beta\cos\phi\,\hatz\nonumber\\
&&-{r\over C}\Bigl[\Delta\Omega
\sin\chi\sin(\phi-\phi_\mu)-\omega\cos\chi\sin\beta\cos\phi \Bigr]
\nonumber\\
&&\times\left[2\sin\chi\cos(\phi-\phi_\mu)\,\hatr+\sin\chi\sin(\phi-\phi_\mu)\,
\hat\phi-\cos\chi\,\hatz\right],
\ea
where
\ba
C &\equiv& 1+3\sin^2\!\chi\cos^2\!(\phi-\phi_\mu),\\
\Delta\Omega &\equiv& \Omega-\omega\cos\beta.
\ea
The drag constant can be written as
\be
\tau_d^{-1}={d^2\mu^2\over m c_A r^6}C
=\tau_{d0}^{-1}C^{1/2},
\label{drag}\ee
where 
\be
\tau_{d0}^{-1}={d^2\mu\over mr^3}({4\pi\rho_0})^{1/2},
\ee
($\rho_0$ is the density of the interblob medium).
The second equality of eq.~(\ref{drag}) is 
valid only when $c_A$ is less than the speed of light.

\subsection{Magnetic Torques}

The magnetic torque on the blob (per unit mass) is given by
\ba
\bN &=& -{r\over\tau_{d0}}C^{1/2}\,\hatr\times\Delta\bV_\perp \nonumber\\
&=&-{r^2\over\tau_{d0}}\Biggl\{
C^{1/2}\left(\Delta\Omega \,\hatz
-\omega \sin\beta\cos\phi\,\hat\phi\right)\nonumber\\
&&-C^{-1/2}\Bigl[\Delta\Omega\,\sin\chi\sin(\phi-\phi_\mu)
-\omega\cos\chi\sin\beta\cos\phi \Bigr]\nonumber\\
&&\times\left[\sin\chi\sin(\phi-\phi_\mu)\,
\hatz+\cos\chi\,\hat\phi\right]\Biggr\}.
\ea
To average over $\phi$, we shall approximate $C^{1/2}$ by $C_1^{1/2}$ and
$C^{-1/2}$ by $C_2^{-1/2}$, where $C_1$ and $C_2$ are constants in the
range of $1-4$. Since the Taylor expansion of $C^{1/2}$ or
$C^{-1/2}$ contains $\cos 2n(\phi-\phi_\mu)$ (where $n$ is an integer),
we can be sure that no new term (with different dependence on the angles)
would appear if the exact expression of $C^{1/2}$ or $C^{-1/2}$ were 
adopted. We find
\ba
\langle\bN\rangle_\phi &=&-{r^2\over\tau_{d0}}C_1^{1/2}
\left(\Delta\Omega\,\hatz
-{1\over 2}\omega \sin\beta\,\haty\right)-{r^2\over 2\tau_{d0}}C_2^{-1/2}
\Biggl[\Delta\Omega
\sin\chi\cos\chi\cos\phi_\mu\,\hatx\nonumber\\
&&+\left(\Delta\Omega
\sin\chi\cos\chi\sin\phi_\mu+\omega\sin\beta\cos^2\!\chi\right)\haty
\nonumber\\
&&-\left(\Delta\Omega\sin^2\!\chi
+\omega\sin\beta\sin\chi\cos\chi\sin\phi_\mu\right)\hatz\Biggr].
\ea
Averaging over the rotation period and using the identities 
(\ref{id1})-(\ref{id3}) we obtain
\ba
\langle\langle\bN\rangle\rangle &=&-{r^2\over\tau_{d0}}C_1^{1/2}
\left(\Delta\Omega\,\hatz
-{1\over 2}\omega \sin\beta\,\haty\right)\nonumber\\
&&-{r^2\over 2\tau_{d0}}C_2^{-1/2}
\Biggl\{\left[\omega\cos^2\!\theta+{1\over 2}\left(\Omega\cos\beta
-\omega\right)\left(3\cos^2\!\theta-1\right)\right]
\sin\beta\,\haty\nonumber\\
&&-\left[\Delta\Omega\sin^2\!\theta+{1\over 2}\Omega\sin^2\!\beta\left(
3\cos^2\!\theta-1\right)\right]\hatz\Biggr\}.
\ea

In the case of $\beta=0$ (i.e., aligned $\hato$ and $\hatl$), the
only nonzero torque is along the $z$-axis:
\be
\langle\langle\bN\rangle\rangle=
-{r^2\over\tau_{d0}}\left(
C_1^{1/2}-{1\over 2}C_2^{-1/2}\sin^2\!\theta\right)\left(\Omega-\omega\right)
\hatz,\quad\qquad\qquad (\beta=0).
\ee
This is simply the magnetic breaking torque which tends to
drive the blob toward corotation with the magnetic field lines. 
For $\beta\neq 0$, there is another torque in the $y$-direction, which
tends to tilt the orbit of the blob. In the absence of other forces,
the tilt angle $\beta$ evloves according to
\be
{d\beta\over dt}={1\over 4\tau_{d0}}\sin\beta\left[C_2^{-1/2}
\left(3\cos^2\!\theta-1\right)\cos\beta
+{\omega\over\Omega}\left(C_2^{-1/2}\sin^2\!\theta-2
C_1^{1/2}\right)\right].
\ee
When $\omega\ll\Omega$, this equation has a similar structure
as eq.~(\ref{dbeta1}). 
Clearly, under certain conditions (when the quantity inside the sqare
bracket is positive), there is an instability
where $\beta$ tends to grow toward the perpendicular state ($\beta=90^\circ$).
The growth time is of order $\tau_{d0}$.

\subsection{Resonances}

The vertical magnetic force (per unit mass) on the blob is given by
\ba
F_z&=&-\tau_{d0}^{-1}C^{1/2}\omega r\sin\beta\cos\phi\nonumber\\
&&-\tau_{d0}^{-1}C^{-1/2}r\cos\chi\Bigl[\Delta\Omega\sin\chi
\sin(\phi-\phi_\mu)-\omega\cos\!\chi\sin\!\beta\cos\phi\Bigr].
\ea
%Following the orbital motion of the blob we have $\phi=\Omega t+\phi_0$.
The equation of motion for the vertical motion is
simply 
\be
{d^2 Z\over dt^2}+\Omega_z^2Z=F_z.
\ee

First consider the case where $\beta=0$. The time-dependence of
the force is as in $C^{-1/2}\sin(\phi-\omega t)$, which can be written as
a sum of $\sin (2n+1)(\phi-\omega t)$. Thus the resonance conditions are
$(2n+1)(\omega-\Omega)=\pm\Omega_z$, or, for $\Omega_z=\Omega$ 
(Keplerian disk):
\be
\omega={2n\over 2n+1}\Omega,\qquad{\rm or}\qquad
\omega={2n+2\over 2n+1}\Omega,\qquad\quad(n=0,1,2,\cdots).
\ee
Thus the resonance ``band'' lies in $\omega/2\le\Omega\le 3\omega/2$
(but excluding $\omega=\Omega$). This was first identified by 
Vietri \& Stella (1998). Note that for $\omega=0$, the resonance 
$\Omega_z=\Omega$ is always satisfied, which implies that the blob will be
driven out of the equatorial plane. 

Now for the general $\beta\neq 0$ cases. If we treat $C$ as being
independent of time, then it is easy to identify the following
resonances: (i) $\Omega_z=\Omega$ (which is always satisfied 
for a Keplerian disk); (ii) $\omega\pm\Omega=\pm\Omega_z$;
(iii) $2\omega\pm\Omega=\pm\Omega_z$. 
To take account of the time-dependence of $C$, we note that
\be 
C=1+3\left[\sin\theta\sin^2\!{\beta\over 2}\cos(\phi+\omega t)
+\sin\theta\cos^2\!{\beta\over 2}\cos(\phi-\omega t)
+\cos\theta\sin\beta\sin\phi\right]^2.
\ee
Thus $C^{1/2}$ or $C^{-1/2}$ can be written as
a sum of $\cos(2n\phi\pm m\omega t)$ and $\sin(2n\phi\pm m\omega t)$
(where $n,m=0,1,2,\cdots$). We can then show that resonances 
occur when $2n\Omega=m\omega$ (for $\Omega_z=\Omega$), i.e., 
all blobs (with different orbital radii) are subjected to resonant 
forcing. We expect that the high-order resonances (those with large $n$
and $m$) are weak, although we have not tried to quantify them. 
The ubiquity of the vertical resonances implies that 
the disk consisting of diamagnetic blobs tends to thicken due to 
the magnetic drag force. 

The radial force (per unit mass) on the blob is given by
\be
F_r=2\tau_{d0}^{-1}C^{-1/2}r\sin\chi\cos(\phi-\phi_\mu)
\Bigl[\Delta\Omega\sin\chi
\sin(\phi-\phi_\mu)-\omega\cos\!\chi\sin\!\beta\cos\phi\Bigr].
\ee
Similar consideration reveals the existence of a large number 
of epicyclic resonances in the disk.

%%%%%%%%%%%%%%%%%%%%%%%%%%%%%%%%%%%%%%%%%%%%%%%%%%%%%%%%%%%%%%%%%%%%%

%%%%%%%%%%%%%%%%%%%%%%%%%%%%%%%%%%%%%%%%%%%%%%%%%%%%%%%%%%%%%%%%%%%%
%\bigskip
\clearpage

%\figcaption{
\begin{figure}
\plotone{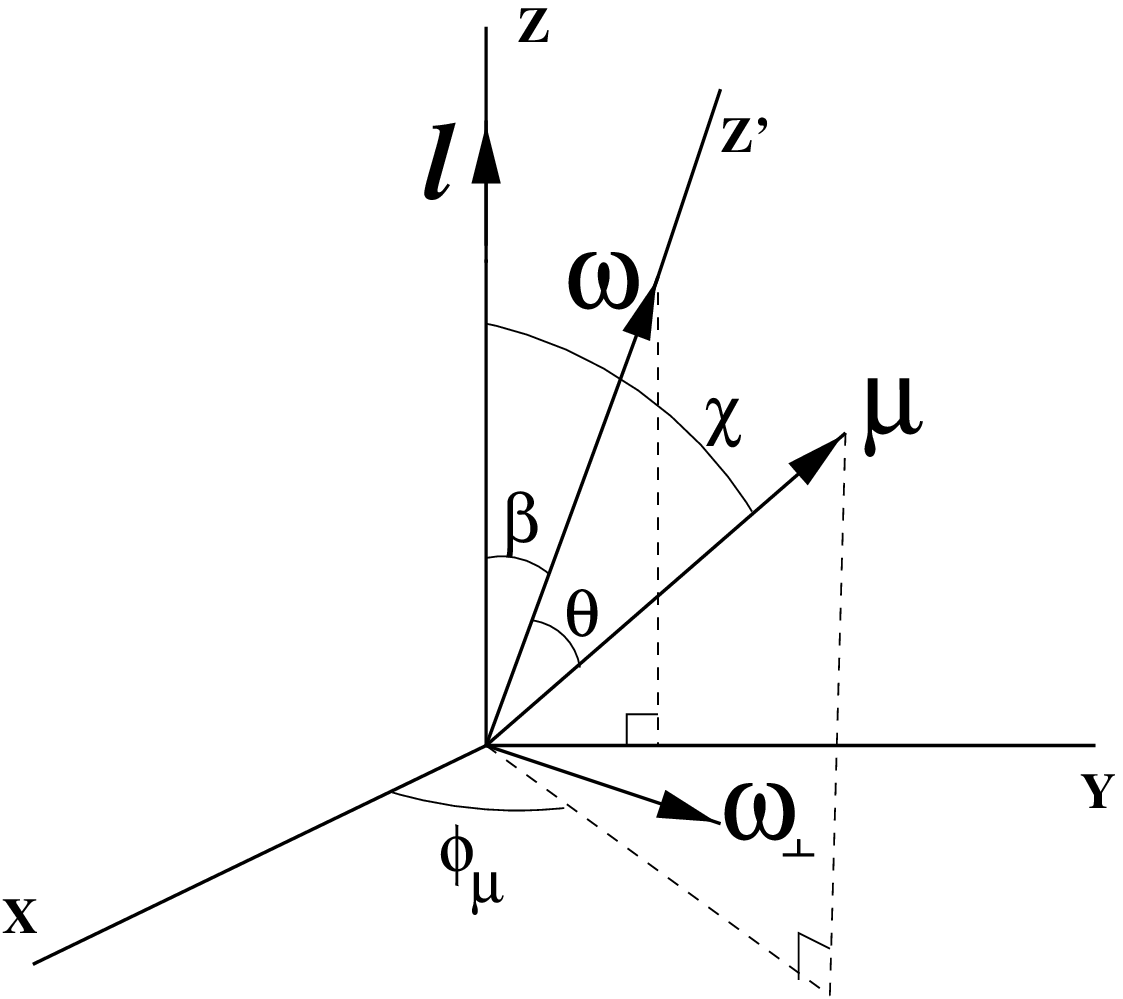}
\caption{
Coordinate system used in the calculation. The $z$-axis is along 
{\boldmath $l$}, the angular momentum of the disk (or a ring of disk).
The angular velocity vector of the star, 
${\mbox{\boldmath $\omega$}}=\omega\hato$ (where $\hato$ is the unit vector) 
is inclined at an angle
$\beta$ with respect to {\boldmath $l$},
and lies in the $yz$-plane. The stellar dipole moment 
${\mbox{\boldmath $\mu$}}=\mu\hat\mu$ (where $\hat\mu$ is the unit vector)
rotates around $\hato$, and
the angle of obliquity is $\theta$. In the cartesian coodinate, we have
$\hat\mu=\sin\chi\cos\phi_\mu\,\hatx+\sin\chi\sin\phi_\mu\,\haty
+\cos\chi\,\hatz$. The axis $z'$ (along the spin axis) is used for studying
disk warping, and ${\mbox{\boldmath $\omega$}}_\perp$ is the unit vector
perpendicular to ${\mbox{\boldmath $\omega$}}$ and lies in the $yz$-plane.
\label{fig1}}
\end{figure}


\begin{thebibliography}{}

\bibitem[]{} 
Agapitou, V., Papaloizou, J.~C.~B., \& Terquem, C. 1997, MNRAS, 292, 631.

\bibitem[]{} 
Alpar, A., Cheng, A.~F., Ruderman, M., \& Shaham, J. 1982, Nature, 
300, 728.

\bibitem[]{} 
Alpar, M.~A., \& Shaham, J. 1985, Nature, 316, 239.

\bibitem[]{} 
Aly, J.~J. 1980, A\&A, 86, 192.
% Exact solution of diamagnetic disk

%\bibitem[]{} 
%Aly, J.~J. 1985, A\&A, 143, 19.

%\bibitem[]{} 
%Aly, J.~J. 1991, ApJ, 375, L61.

\bibitem[]{} 
Aly, J.~J., \& Kuijpers, J. 1990, A\&A, 227, 473.

\bibitem[]{} 
Anzer, U., \& B\"orner, G. 1980, A\&A, 83, 133.

\bibitem[]{} 
Anzer, U., \& B\"orner, G. 1983, A\&A, 122, 73.

\bibitem[]{} 
Armitage, P.~J., \& Clarke, C.~J. 1996, MNRAS, 280, 458.
%Magnetic Breaking of T Tauri Stars

\bibitem[]{} 
Arons, J. 1987, in ``The Origin and Evolution of Neutron Stars'' (IAU
Symp.~No.~125), ed. D.~J. Helfand \& J.-H. Huang (D. Reidel Pub.:
Dordrecht).

\bibitem[]{} 
Arons, J. 1993, ApJ, 408, 160.

\bibitem[]{} 
Arons, J., \& Lea, S.~M. 1980, ApJ, 235, 1016.

\bibitem[]{} 
Bardeen, J.~M., \& Petterson, J.~A. 1975, ApJ, 195, L65.

\bibitem[]{} 
Basri, G, Marcy, G.~W., \& Valenti, J.~A. 1992, ApJ, 390, 622.

\bibitem[]{} 
Bertout, C. 1989, ARA\&A, 27, 375.

\bibitem[]{} 
Bhattacharya, D. 1995, in ``X-ray Binaries'', ed. W.H.G. Lewin, 
J. van Paradijs \& E.P.J. van den Heuvel (Cambridge Univ. Press).

\bibitem[]{} 
Bildsten, L., et al.~1997, ApJS, 113, 367.

\bibitem[]{} 
Blandford, R.~D. 1989, in Theory of Accretion Disks, ed. F. Meyer et al.
(Kluwer Academic Pub.: Dordrecht).

\bibitem[]{} 
Bouvier, J., et al. 1995, A\&A, 299, 89.

\bibitem[]{} 
Burns, J.~A., Schaffer, L.~E., Greenberg, R.~J., \& Showalter, M.~R.
1985, Nature, 316, 115.
% Lorentz resonances and the structure of the Jovian ring

\bibitem[]{} 
Cameron, A.~C., \& Campbell, C.~G. 1993, A\&A, 274, 309.
%Rotational Evolution of magnetic T Tauri stars with accretion disks

\bibitem[]{} 
Campbell, C.~G. 1997, Magnetohydrodynamics in Binary Stars
(Kluwer Academic Pub.: Dordrecht).

\bibitem[]{} 
Chakrabarty, D., et al.~1997, ApJ, 481, L101.
%Torque and L in GX 1+4

\bibitem[]{} 
Deeter, J.~E., et al. 1998, ApJ, 502, 802.
% The 35 day evolution of the Her X-1 pulse profile

\bibitem[]{} 
Drell, S.~D., Foley, H.~M., \& Ruderman, M.~A. 1965, J. Geophys. Res.,
70, 3131. 

\bibitem[]{} 
Ford, E.~C., \& van der Klis, M. 1998, 506, L39.
% (astro-ph/9808148).
%Strong Correlation Between Noise Features at Low Frequency and the Kilohertz
%QPOs in the X-Ray Binary 4U 1728-34

\bibitem[]{} 
Frank, J., King, A., \& Raine, D. 1992, Accretion Power in Astrophysics
(Cambridge Univ. Press). 

\bibitem[]{} 
Ghosh, P., \& Lamb, F.~K. 1979a, ApJ, 232, 259.

\bibitem[]{} 
Ghosh, P., \& Lamb, F.~K. 1979b, ApJ, 234, 296.

\bibitem[]{} 
Ghosh, P., \& Lamb, F.~K. 1992, in X-ray Binaries and Recycled Pulsars, ed.
E.P.J. van den Heuval and S.A. Rappaport (Dordrecht: Kluwer).

\bibitem[]{} 
Goodson, A.~P., Winglee, R.~M., \& B\"ohm, K.-H. 1997, ApJ, 489, 199.
% Time-dependent accretion by magnetic young stellar objects as a launching
% mechanism for stellar jets

\bibitem[]{} 
Hartmann, L. 1998, Accretion Processes in Star Formation (Cambridge
Univ. Press).

\bibitem[]{} 
Hasinger, G., \& van der Klis, M. 1989, A\&A, 225, 79.

\bibitem[]{} 
Hayashi, M.~R., Shibata, K., \& Matsumoto, R. 1996, ApJ, 468, L37.

\bibitem[]{} 
Ipser, J.~R. 1996, ApJ, 458, 508.

\bibitem[]{} 
Katz, J.~I. 1973, Nature Phys. Sci., 246, 87.

\bibitem[]{} 
King. A.~R. 1993, MNRAS, 261, 144.

\bibitem[]{} 
K\"onigl, A. 1991, ApJ, 370, L39.

%\bibitem[]{} 
%Kuburaki, O. 1986, MNRAS, 220, 321.
%Ghosh-Lamb type: in detail

\bibitem[]{} 
Kumar, S., \& Pringle, J.~E. 1992, MNRAS, 258, 811.

\bibitem[]{} 
Kundt, W., \& Robnik, M. 1980, A\&A, 91, 305.

\bibitem[]{} 
Lai, D. 1998, ApJ, 502, 721.

\bibitem[]{} 
Lai, D., Lovelace, R.~V.~E., \& Wasserman, I. 1999, submitted to ApJ
(astro-ph/9904111).

\bibitem[]{} 
Lamb, F.~K., Pethick, C.~J., \& Pines, D. 1973, ApJ, 184, 271.

\bibitem[]{} 
Lamb, F.~K., Shibazaki, N., Alpar, M.~A., \& Shaham, J. 1985, Nature, 317, 681.

\bibitem[]{} 
Lang, F.~L., et al. 1981, ApJ, 246, L21.

\bibitem[]{} 
Li, J., Wickramasinge, D.~T., \& R\"udiger, G. 1996, ApJ, 469, 765.
%Magnetized Accretion and Funnel Flow

\bibitem[]{} 
Lipunov, V.~M., Sem\"enov, E.~S., \& Shakura, N.~I. 1981, Sov. Astron., 25,
439.

\bibitem[]{} 
Lipunov, V.~M., \& Shakura, N.~I. 1980, Sov. Astron. Lett., 6, 14.

%\bibitem[]{} 
%Lovelace, R.~V.~E., Wang, J.~C.~L., \& Sulkanen, M.~E. 1987, ApJ, 315, 504.

\bibitem[]{} 
Lovelace, R.~V.~E., Romanova, M.~M., \& Bisnovatyi-Kogan, G.~S. 1995,
MNRAS, 275, 244.

\bibitem[]{} 
Lovelace, R.~V.~E., Romanova, M.~M., \& Bisnovatyi-Kogan, G.~S. 1998,
ApJ, in press (astro-ph/9811369).
% Magnetic Propeller Outflows

\bibitem[]{} 
Lynden-Bell, D., \& Boily, C. 1994, MNRAS, 267, 146.

\bibitem[]{} 
Maloney, P.~R., \& Begelman, M.~C. 1997, ApJ, 491, L43.

\bibitem[]{} 
Margon, B. 1984, ARA\&A, 22, 507.

\bibitem[]{} 
Markovi\'c, D., \& Lamb, F.~K. 1998, ApJ, 507, 316.

\bibitem[]{} 
M\'endez, M., et al. 1998a, ApJ, 494, L65.
% (astro-ph/9807281
% kHz QPO peak separation is not constant in the Atoll source 4U 1608-52.

\bibitem[]{} 
M\'endez, M., van der Klis, M., \& van Paradijs, J. 1998b, ApJ, 
505, L23. 
%(astro-ph/9808281).
% The kHz QPO peak separation is not equal to half the burst oscillation
% frequency in 4U 1636-53.

\bibitem[]{} 
Miller, M.~C., Lamb, F.~K., \& Psaltis, D. 1998, ApJ, 508, 791.

\bibitem[]{} 
Miller, K.~A., \& Stone, J.~M. 1997, ApJ, 489, 890.

\bibitem[]{} 
Morsink, S.~M., \& Stella, L. 1999, ApJ, 513, 827.
%astro-ph/9808227.
%Relativistic Precession around Rotating Neutron Stars: Effects Due to Frame
%Dragging and Stellar Oblateness

\bibitem[]{} 
Nelson, R.~W., et al.~1997, ApJ, 488, L117.

\bibitem[]{} 
Newman, W.~I., Newman, A.~L., \& Lovelace, R.~V.~E. 1992, ApJ, 392, 622.

\bibitem[]{} 
Novikov, I.~D., \& Thorne, K.~S. 1973, in Black Holes, ed.
C. DeWitt and B. DeWitt (Gordon and Breach: New York).

\bibitem[]{} 
Ogilvie, G.~I. 1999, MNRAS, 304, 557.
%(astro-ph/9812073).
%The non-linear fluid dynamics of a warped accretion disc

\bibitem[]{} 
Papaloizou, J.~C., \& Pringle, J.~E. 1983, MNRAS, 202, 1181.

\bibitem[]{} 
Park, S.~J., \& Vishniac, E.~T. 1996, ApJ, 471, 158.

\bibitem[]{} 
Phinney, E.~S., \& Kulkarni, S.~R. 1994, ARA\&A, 32, 591.

\bibitem[]{} 
Priedhorsky, W.~C., \& Holt, S.~S. 1987, Space Sci. Rev., 45, 291.

\bibitem[]{} 
Pringle, J.~E. 1992, MNRAS, 258, 811.

\bibitem[]{} 
Pringle, J.~E. 1996, MNRAS, 281, 857.

\bibitem[]{} 
Pringle, J.~E., \& Rees, M.~J. 1972, A\&A, 21, 1.

\bibitem[]{} 
Psaltis, D., et al. 1999, submitted to ApJ.
% On the magnetosphere beat-freq. and L-T interpretations of
% HBOs in Z sources

\bibitem[]{} 
Riffert, H. 1980, Astro. Space. Sci., 71, 195.

\bibitem[]{} 
Schaffer, L., \& Burns, J.~A. 1992, ICARUS, 96, 65.

\bibitem[]{} 
Shakura, N.~I., \& Sunyaev, R.~A. 1973, A\&A, 24, 337.

\bibitem[]{} 
Sheffer, E.~K., et al. 1992, Sov. Astron., 36, 41.

\bibitem[]{} 
Shu, F.~H., et al.~1994, ApJ, 429, 781.

\bibitem[]{} 
Spruit, H.~C., \& Taam, R.~E. 1990, A\&A, 229, 475.

\bibitem[]{} 
Spruit, H.~C., \& Taam, R.~E. 1993, ApJ, 402, 593.

\bibitem[]{} 
Scheuer, P.~A.~G., \& Feiler, R. 1996, MNRAS, 282, 291.

\bibitem[]{} 
Stella, L., \& Vietri, M. 1998, ApJ, 492, L59.
% (astro-ph/9709085).

\bibitem[]{} 
Stella, L., \& Vietri, M. 1999, Phys. Rev. Lett., in press
(astro-ph/9812124).

\bibitem[]{} 
Stone, J.~M., \& Norman, M.~L. 1994, ApJ, 433, 746.

%\bibitem[]{} 
%Sturrock, P.~A. 1991, ApJ, 380, 655.

\bibitem[]{} 
Sturrock, P.~A., \& Barnes, C. 1972, ApJ, 176, 31.
% Reconnection limits the azimuthal pitch to 1.

\bibitem[]{} 
Tananbaum, H., et al. 1972, ApJ, 174, L143.

\bibitem[]{} 
Torkelsson, U. 1998, astro-ph/9803068.

\bibitem[]{} 
Toropin, Y.~M., et al. 1999, ApJ, in press.
% Spherical accretion on to a magnetic dipole

\bibitem[]{} 
van Ballegooijen, A.~A. 1994, Space Science Rev., 68, 299.

\bibitem[]{} 
Van der Klis, M. 1995, in ``X-ray Binaries'', ed. W.H.G. Lewin, 
J. van Paradijs \& E.P.J. van den Heuvel (Cambridge Univ. Press).

\bibitem[]{} 
Van der Klis, M. 1998a, in ``The Many Faces of Neutron Stars'' (Proc.
NATO ASI) (astro-ph/9710016).
%the Proceedings of the Wise Observatory 25th
%Anniversary Symposium "Astronomical Time Series" (astro-ph/9704272).

\bibitem[]{} 
Van der Klis, M. 1998b, astro-ph/9812395.

\bibitem[]{} 
Van der Klis, M., et al. 1985, Nature, 316, 225.
% Discovery of HBO

\bibitem[]{} 
Van der Klis, M., Wijnands, R.~A., Horne, K., \& Chen, W. 1997, ApJ,
481, L97.
% Sco X-1: separation frequency of the two peaks is not constant

\bibitem[]{} 
Van Kerkwijk, M.~H., et al. 1998, ApJ, 499, L27.
% Warped Disks as a possible origin of torque reversals in accretion-powered
% Pulsars

\bibitem[]{} 
Vietri, M., \& Stella, L. 1998, ApJ, 503, 350.
%in press (astro-ph/9803089).

%\bibitem[]{} 
%Wang, J.~C.~L., Sulkanen, M.~E., \& Lovelace, R.~V.~E. 1990, ApJ, 355, 38.

\bibitem[]{} 
Wang, Y.-M. 1987, A\&A, 183, 257.

\bibitem[]{} 
Wang, Y.-M. 1995, ApJ, 449, L153.

\bibitem[]{} 
Wijers, R.~A.~M.~J., \& Pringle, J.~E. 1998, astro-ph/9811056.

\bibitem[]{} 
Yi, I. 1995, ApJ, 442, 768.
% Magnetized accretion and the spin evolution of classical T Tauri stars

\bibitem[]{} 
Yi, I., \& Kenyon, S.~J. 1997, ApJ, 477, 379.
%A Laboratory for Magnetized Accretion Disk Model: Ultraviolet and X-Ray
%Emission from Cataclysmic Variable GK Persei

\bibitem[]{} 
Yi, I., Wheeler, C., \& Vishniac, E.~T. 1997, ApJ, 481, L51.

\end{thebibliography}
\end{document}